# Identification of Physical Processes and Unknown Parameters of 3D Groundwater Contaminant Problems via Theory-guided U-net

Tianhao He[a], Haibin Chang[b], and Dongxiao Zhang[c,d*]

[a]College of Engineering, Peking University, Beijing 100871, P. R. China
[b]School of Energy and Mining Engineering, China University of Mining and Technology (Beijing), Beijing 100083, P. R. China
[c]School of Environmental Science and Engineering, Southern University of Science and Technology, Shenzhen 518055, P. R. China
[d] Department of Mathematics and Theories, Peng Cheng Laboratory, Shenzhen 518000, P. R. China
[*]Corresponding author: E-mail address: zhangdx@sustech.edu.cn (Dongxiao Zhang)

**Abstract**

Identification of unknown physical processes and parameters of groundwater contaminant sources is a challenging task due to their ill-posed and non-unique nature. Numerous works have focused on determining nonlinear physical processes through model selection methods. However, identifying corresponding nonlinear systems for different physical phenomena using numerical methods can be computationally prohibitive. With the advent of machine learning (ML) algorithms, more efficient surrogate models based on neural networks (NNs) have been developed in various disciplines. In this work, a theory-guided U-net (TgU-net) framework is proposed for surrogate modeling of three-dimensional (3D) groundwater contaminant problems in order to efficiently elucidate their involved processes and unknown parameters. In TgU-net, the underlying governing equations are embedded into the loss function of U-net as soft constraints. For the considered groundwater contaminant problem, sorption is considered to be a potential process of an uncertain type, and three equilibrium sorption isotherm types (i.e., linear, Freundlich, and Langmuir) are considered. Different from traditional approaches in which one model corresponds to one equation, these three sorption types are modeled through only one TgU-net surrogate. The three mentioned sorption terms are integrated into one equation by assigning indicators. Accurate predictions illustrate the satisfactory generalizability and extrapolability of the constructed TgU-net. Furthermore, based on the constructed TgU-net surrogate, a data assimilation method is employed to identify the physical process and parameters simultaneously. The convergence of indicators demonstrates the validity of the proposed method. The influence of sparsity-promoting techniques, data noise, and quantity of observation information is also explored. This work shows the possibility of governing equation discovery of physical problems that contain multiple and even uncertain processes by using deep learning and data assimilation methods.

## 1. Introduction

The physical process of contaminant transport problems usually consists of advection, dispersion, and reaction (Fetter, 1999; Zheng and Wang, 1999), which may be unclear in a contaminant field. Identification of the physical process and characteristics of a contaminant source is important for modeling and predicting the further concentration distribution. This can be regarded as a high-dimensional inverse problem (Zhang et al., 2015; Xu and Gómez-Hernández, 2018). However, complex heterogeneous geological conditions cause the ill-posed and non-unique nature of contaminant source identification, which makes it especially challenging to simultaneously determine the physical process and unknown parameters.

In current works, model selection and symbolic regression methods have become the mainstream of discovering physical equations. The core idea of model selection methods is building a large set of potential candidate models, and then selecting the model with the maximum probability or information criteria (IC) as the proper one to describe the physical process. Model selection methods have been widely applied to decrease model structure uncertainty. The Akaike, Bayesian, and Kashyap information criteria (AIC, BIC, and KIC) (Kuha, 2004; Ye et al., 2008) are usually employed to assess likelihood in model selection methods. Schoeniger et al. (2014) used Monte Carlo (MC) integration methods as a reference and compared nine ways to implement model selection. The results demonstrated that the chosen models are heavily biased by ICs when applied to nonlinear systems, yielding inaccurate predictions. If the parameters are high-dimensional or the physical process is strongly nonlinear, model selection methods based on ICs may be computationally prohibitive. Cao et al. (2019) utilized a Bayesian model selection method to implement groundwater contaminant source identification. They applied the POLYCHORD algorithm to estimate the marginal likelihood and posterior distribution of potential models. POLYCHORD is a novel nested sampling system with strong search ability for high-dimensional distribution. It can calculate the marginal likelihood and posterior distribution of parameters simultaneously. The model with the highest marginal likelihood can best describe the physical phenomenon. The model selection methods based on ICs require an artificially established model set, which may be greatly influenced by subjectivity. In contrast, symbolic regression methods can automatically combine derivatives from a candidate differential terms library, and the combination of differential terms that can best fit observation data is regarded as the governing equation. In order to simplify the discovered equations to the greatest extent under the premise of maintaining accuracy, some sparsity-promoting techniques can be employed, such as the sequential thresholded least-squares (LASSO) algorithm. A penalty function is built in LASSO to compress variables sparsely, and make some regression coefficients become 0 to achieve the purpose of variable selection (Brunton et al., 2016; Schaeffer et al., 2017). Mangan et al. (2017) proposed to trace evaluation criteria, such as AIC or BIC, in symbolic regression, which can automatically search for the optimal model that provides the most information. However, traditional symbolic regression methods are sensitive to noise. In addition, some novel algorithms based on candidate libraries have been proposed. Ying et al. (2017) combined three biotransformation processes of a dynamic contaminant problem by defining a generalized rate equation. The model selection problem and parameter estimation are solved by the Bayesian inference method. Chang and Zhang (2019) proposed to utilize a data assimilation method to discover the proper terms from a candidate library. Data assimilation methods can resist a certain level of noise. They firstly established a candidate library, consisting of advection, dispersion, and two sorption terms of a contaminant problem. The searched models were trained in data-driven form. Then, the iterative ensemble smoother (IES) method was employed to estimate the sorption type. The results demonstrate the capability of discovering equations via data assimilation methods.

The traditional methods mentioned above are usually time-consuming due to the requisite large number of model executions. It is also difficult to calculate information criteria if the system is strongly nonlinear. Consequently, the accuracy of the searched model may be further influenced (Troldborg et al., 2010; Schoeniger et al., 2014). A fast and efficient method for solving forward problems is needed. In recent years, with the development of artificial intelligence (AI), deep learning algorithms, especially neural networks (NNs), have been widely applied to construct surrogate models to facilitate calculation speed. Mo et al. (2019a; 2019b) proposed to construct encoder-decoder surrogate models with 686 uncertain input parameters of contaminant source. Results demonstrate the strong fitting ability, extrapolability, and high dimensional problem processing ability of neural network surrogate models. They further employed the iterative local updating ensemble smoother (ILUES), a data assimilation method, to estimate source strength. The convergence of the estimated parameters to the true values shows the validity of established surrogate models. Srivastava and Singh (2015) constructed several artificial neural network (ANN) surrogate models that mapped obvservation data to contaminant source parameters directly. These models were tested, and the model with the best performance was chosen. They found that the difficulty of identifying unknown parameters increases with the decrease of information. Moreover, ML algorithms, such as the genetic algorithm (GA), have attracted great attention regarding searching for the optimal equation. Different from other algorithms that identify the optimal solution, GA has no derivation and no restriction on the continuity of function. Specifically, it adopts the probabilistic optimization method. GA can automatically obtain and guide the optimized search space without definite rules, and can adaptively adjust the search direction. Xu et al. (2020) proposed a novel DLGA-PDE framework for discovering partial differential equations (PDEs) from an established candidate library. They employed ANNs to approximate responses and utilized GA to search the optimal matching equation. The DLGA-PDE method is robust in both equation discovery and noise resistance.

In the present work, we aim to construct highly efficient ML-based surrogate models, and then utilize a data assimilation method to identify the physical process and unknown parameters of a contaminant source simultaneously. NNs are employed in this work for surrogate modeling, and it is well known that a pure data-driven training scheme cannot guarantee the generalizability and extrapolability of NNs. On the other hand, data acquisition is usually costly, which may limit the application of NNs that require a large volume of training data. Recent investigations have focused more on incorporating underlying theoretical information or equations into NN modeling to improve their predictive accuracy. Raissi et al. (2019) proposed to incorporate the residual of governing equations into the loss function of NNs, i.e., physics-informed neural networks (PINNs). Yang et al. (2020) developed physics-informed generative adversarial networks (PIGANs). They calculated the differential terms via different networks, and combined them in the structure of PIGAN to express the governing equations. On the basis of PINN, Wang et al. (2020a; 2021a) proposed a more general theory-guided nerual networks (TgNNs) framework, in which more factors, such as engineer experience and control conditions, are considered as soft constraints and added in the loss function of TgNNs. Xu et al. (2021) calculated PDEs in weak form to further improve the accuracy of TgNNs. In order to process image-like data more conveniently, Wang et al. (2021b) applied theory-guidance to an encoder-decoder structure, and proposed theory-guided convolutional neural networks (TgCNNs) for well placement optimization problems. He et al. (2021) developed theory-guided full convolutional neural netwroks (TgFCNNs) for problems with high localized features, such as contaminant transport. The structure of TgFCNNs is simpler and easier to build compared to TgCNN because only several convolutional layers are considered. It is worth noting that the works metioned above focus on 2D problems (Wu et al., 2019; Zhou and Tartakovsky, 2021). The difficulty of fitting observation data increases with problem dimension increase, and more efficient models are needed for high-dimensional problems. To this end, a theory-guided U-net framework (TgU-net) is proposed to precisely deal with 3D contaminant transport problems in this work. U-net was originally proposed for biomedical image segmentation (Ronneberger et al., 2015). It is so-named due to its distinctive U-shaped structure. U-net possesses a similar structure to encoder-decoder nets. Traditional encoder-decoders only use the deepest extracted feature, while U-net adopts the skip-connection strategy to make full use of features obtained by all enocders. Moreover, it connects corresponding symmetrical layers of the encoder-decoder to make up the lost information in the encoding stage. In addition, more high-resolution information can be fully utilized in the up-sampling process. U-net-based algorithms have been broadly applied in computational fluid dynamics (CFD) and hydrological projects (Chen et al., 2020; Chun-Yu et al., 2021; Jiang et al., 2021; Le and Ooi, 2021; Lee and Park, 2021; Tang et al., 2020; Wang et al., 2020b).

  Traditionally, one surrogate may only correspond to one physical problem with a fixed model; whereas, the proposed TgU-net can deal with problems with an uncertain process that can be described by multiple potential models. In this work, a 3D contaminant problem is to be solved. Three sorption types (i.e., linear, Freundlich, and Langmuir) are considered, and they are integrated into one equation by assigning indicators. The residual of this comprehensive equation is embedded into the loss function to guide the updating direction of U-net parameters. In order to perform the theory-guided strategy, two TgU-net surrogate models are constructed to predict the flow field and the contaminant field, respectively. It is worth mentioning that there is no need to parameterize conductivity fields due to the nature of convolutional neural networks (CNNs). Furthermore, the iterative ensemble smoother (IES), a data assimilation method, is employed to identify the physical process and parameters of the contaminant source based on the constructed surrogate. The results demonstrate that the TgU-net framework can capture the governing equation satisfactorily. The extrapolability and generalizability of TgU-net are also tested, and the results show that TgU-net can still achieve satisfactory predictive capability when faced with unseen configurations, such as conductivity fields with higher means or variance. The influence of sparsity-promoting techniques, data noise, quality of information, and nonlinearity of equations on the results of data assimilation are discussed.

The remainder of this paper proceeds as follows. The influence of model uncertainty on inverse problems, model selection, symbolic regression, and the purpose of this work are introduced in section 1. This is followed by the presentation of governing equations, the structure of TgU-net, and the IES method in section 2. In section 3, we elaborate on the performance of TgU-net and the results of data assimilation. Finally, the conclusion and discussion are presented in section 4.

## 2. Methodology

In this section, we firstly outline the governing equations of the contaminant convection-dispersion problem. Then, the structure of TgU-net is presented. Finally, the IES method is briefly introduced.

### 2.1 Governing equations

In this work, a 3D contaminant transport problem is considered. The flow field is the prerequisite for the convection of a contaminant. The groundwater flow differential equation is used to express the relationship between the hydraulic conductivity field and the flow field:

$$\frac{\partial}{\partial x_i}\left(K_i \frac{\partial h}{\partial x_i}\right) + q_s = S_s \frac{\partial h}{\partial t} \qquad (1)$$

where $K_i$ denotes the hydraulic conductivity field, a parameter describing formation heterogeneity; $h$ denotes the hydraulic head; $x_i$ denotes the distance along the respective Cartesian coordinate axis (i.e., x, y, and z); $q_s$ is the fluid sink (negative) or source (positive) term; and $S_s$ denotes the specific storage. Herein, a steady state flow field is considered, which means that the right-hand side of **Eq. (1)** is 0.

The contaminant convection-dispersion equation can be expressed as follows:

$$\frac{\partial(\theta C)}{\partial t} = \frac{\partial}{\partial x_i}\left(\theta D_{i,j} \frac{\partial C}{\partial x_j}\right) - \frac{\partial}{\partial x_i}(\theta v_i C) + q_s C_s + \sum R \qquad (2)$$

where $\theta$ denotes the porosity of the subsurface medium; $C$ denotes the solute concentration of contaminant; $D_{i,j}$ denotes the hydrodynamic dispersion coefficient tensors; $v_i$ is the seepage, as calculated in **Eq. (3)**; $C_s$ is the concentration of $q_s$; and $\sum R$ denotes reaction terms, such as chemical reaction or sorption.

$$v_i = -\frac{K_i}{\theta} \frac{\partial h}{\partial x_i} \qquad (3)$$

The hydrodynamic dispersion coefficient tensors $D_{i,j}$ in 3D problems can be obtained as follows:

$$\begin{aligned}
D_{XX} &= \alpha_L \frac{v_x^2}{|v|} + \alpha_{TH} \frac{v_y^2}{|v|} + \alpha_{TV} \frac{v_z^2}{|v|} + D^* \\
D_{YY} &= \alpha_L \frac{v_y^2}{|v|} + \alpha_{TH} \frac{v_x^2}{|v|} + \alpha_{TV} \frac{v_z^2}{|v|} + D^* \\
D_{ZZ} &= \alpha_L \frac{v_z^2}{|v|} + \alpha_{TV} \frac{v_x^2}{|v|} + \alpha_{TV} \frac{v_y^2}{|v|} + D^* \\
D_{XY} &= D_{YX} = (\alpha_L - \alpha_{TH}) \frac{v_x v_y}{|v|} \\
D_{XZ} &= D_{ZX} = (\alpha_L - \alpha_{TV}) \frac{v_x v_z}{|v|} \\
D_{YZ} &= D_{ZY} = (\alpha_L - \alpha_{TV}) \frac{v_y v_z}{|v|} \\
|v| &= \sqrt{v_x^2 + v_y^2 + v_z^2}
\end{aligned} \tag{4}$$

where $\alpha_L$, $\alpha_{TH}$, and $\alpha_{TV}$ denote longitudinal, horizontal transverse, and vertical transverse dispersivity, respectively; and $D^*$ denotes the effective molecular diffusion coefficient.

Reaction term $\sum R$ consists of chemical reaction, sorption, etc., which may occur in the formation to retard the transport of contaminant. In this work, sorption is considered to be a potential process of the contaminant transport problem, and three equilibrium sorption isotherm types are considered, i.e., linear, Freundlich, and Langmuir sorption. It is assumed that the concentration of compounds in the solution is in equilibrium with that adsorbed into the porous medium under constant temperature.

The sorption term in the contaminant convection-dispersion equation is given as follows:

$$\sum R = -\rho_d \frac{\partial \bar{C}}{\partial t} \tag{5}$$

where $\rho_d$ denotes the bulk density of aquifer; and $\bar{C}$ denotes the amount of solute absorbed by solid per unit weight. In equilibrium linear, Freundlich, and Langmuir sorption isotherm, $\bar{C}$ can be calculated as **Eq. (6-8)**, respectively:

$$\bar{C} = K_d C \tag{6}$$

$$\bar{C} = K_f C^a \tag{7}$$

$$\bar{C} = \frac{K_l \bar{S} C}{1 + K_l C} \tag{8}$$

where $K_d$, $K_f$, and $K_l$ denote the distribution coefficient, Freundlich, and Langmuir equilibrium constant, respectively; $a$ is the Freundlich exponent; and $\bar{S}$ is the total concentration of the sorption site available (Zheng and Wang, 1999).

## 2.2 Theory-guided U-net

Contaminant transport problems usually contain a lot of uncertainty. For instance, the occurred process of a field site, the characteristics of the contaminant source, the spatial distribution of the formation property field, etc., may be unclear. As a consequence, inverse modeling is usually required. Inverse modeling is well known to be time-consuming, and building a surrogate model of the considered physical problem is an effective way to speed-up this process. In this work, U-net is employed for surrogate modeling of the contaminant transport problem. U-net is a derivation of the full convolutional neural network and encoder-decoder structure (Shelhamer et al., 2017). There is no need to parameterize 2D or 3D data (e.g., hydraulic conductivity fields) prior to being input, and it can realize the mapping relationship from image to image. The encoder of U-net can compress high dimensional input into low dimensional features step by step. This can be regarded as a down-sampling process. The decoder, however, is the opposite of this. The deepest hidden layer contains the most effective extracted features, which can be decompressed to generate the target image through up-sampling. However, some information may be lost in the up-sampling process, and traditional encoder-decoder neural networks ignore this detail. The advantage of U-net is transferring extracted features from each encoder layer to the symmetric decoder through the skip-connection method, to make up the lost information in the up-sampling process. Information details of the original image can be captured as comprehensively as possible, and accuracy can be further improved (Tang et al., 2018). In recent years, derivations of U-net have been widely developed and applied (Dolz et al., 2019; Newell et al., 2016). The structure of U-net used in this work is shown in **Fig. 1**. This employed U-net consists of one input, output and latent layer, as well as seven encoder-decoder layers. Convolutional and deconvolutional kernel sizes are both (2, 4, 6) for 3D data. The stride is set to be 1, and the padding is set to be 0.

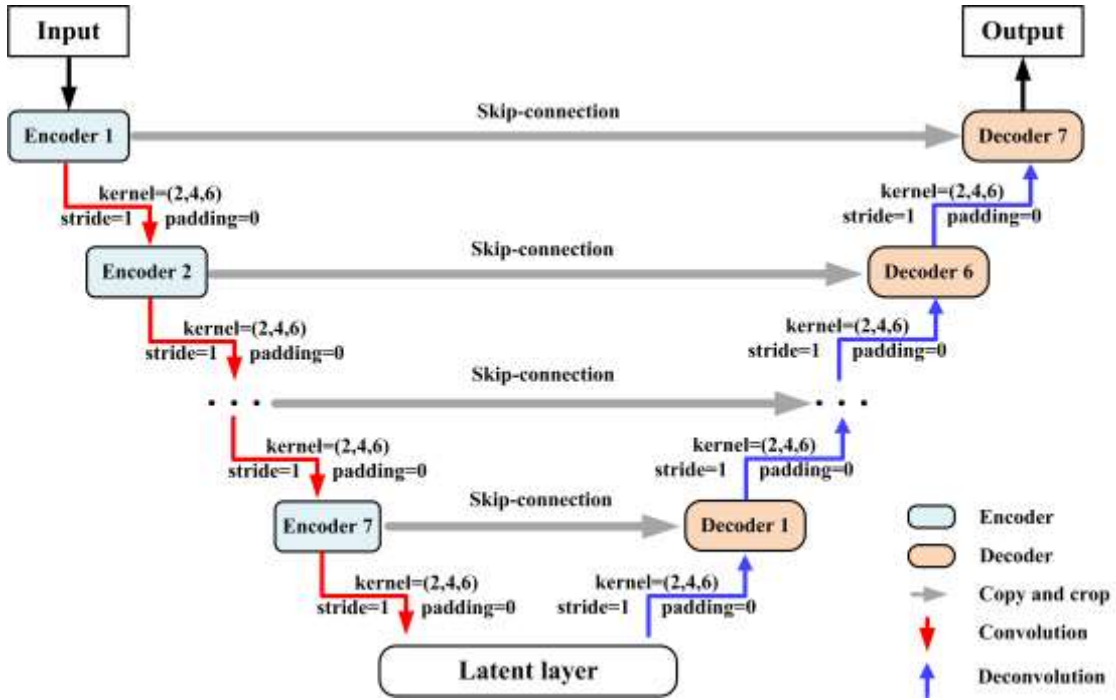

**Figure 1.** Structure of the employed U-net.

Incorporating underlying governing equations into the loss function can make the prediction of neural networks (NNs) consistent with physical laws. In order to realize theory-guided training, two surrogate models should be constructed to fit **Eq. (1)** and **Eq. (2)**, respectively, i.e., TgU-net-h and TgU-net-C. Predictions should be substituted into the equations to calculate the residuals of **Eq. (1)** and **Eq. (2)** in the form of mean square error (MSE), as shown in **Eq. (9)** and **Eq. (10)**, respectively. The seepage $v_i$ can be calculated through the predicted hydraulic head field according to **Eq. (3)**, and be further inputted into the surrogate model for contaminant to calculate **Eq. (10)**:

$$MSE_{pde1} = \frac{1}{N_1}\sum_{n=1}^{N_1}\left(\frac{\partial}{\partial x_i}\left(K_i\frac{\partial \hat{h}_n}{\partial x_i}\right) + q_s - S_s\frac{\partial \hat{h}_n}{\partial t}\right)^2 \quad (9)$$

$$MSE_{pde2} = \frac{1}{N_2}\sum_{n=1}^{N_2}\left(\frac{\partial(\theta\hat{C}_n)}{\partial t} - \frac{\partial}{\partial x_i}\left(\theta D_{i,j}\frac{\partial \hat{C}_n}{\partial x_j}\right) + \frac{\partial}{\partial x_i}\left(\theta v_i\hat{C}_n\right) - q_sC_s - \sum R\right)^2 \quad (10)$$

where $\hat{h}$ and $\hat{C}$ denote the predicted hydraulic head field and contaminant field from corresponding surrogate models; and $N_1$ and $N_2$ are the number of collocation points for the hydraulic head field and contaminant field, respectively. Here, collocation points are the spatial-temporal locations for enforcing PDE constraints, which can be randomly sampled. Herein, $\sum R$ can be presented by the combination of the mentioned three sorption types (i.e., equilibrium linear, Freundlich, and Langmuir sorption isotherm) by assigning indicators, as shown as follows:

$$\sum R = -\rho_d\frac{\partial}{\partial t}\left(\tau_1\gamma_1\hat{C}_n + \tau_2\gamma_1\hat{C}_n^{\gamma_2} + \tau_3\frac{\gamma_1\gamma_2\hat{C}_n}{1+\gamma_1\hat{C}_n}\right) \quad (11)$$

where $\tau_i$ ($i$ = 1, 2, 3) are the indicators that decide the sorption type, taking values of 0 or 1. The sum of $\tau_i$ is also not more than 1. For instance, if the equilibrium Freundlich sorption isotherm is the reaction term of one hydrodynamic contaminant problem, $\tau_1 = \tau_3 = 0$ and $\tau_2 = 1$. If $\tau_1 = \tau_2 = \tau_3 = 0$, there is no sorption. Therefore, we can ascertain sorption type during contaminant transport by estimating values of indicators with the assistance of data assimilation methods. $\gamma_1$ is the distribution coefficient $K_d$ (when $\tau_1 = 1$), or the Freundlich equilibrium constant $K_f$ (when $\tau_2 = 1$), or the Langmuir equilibrium constant $K_l$ (when $\tau_3 = 1$). $\gamma_2$ denotes the Freundlich exponent $a$ (when $\tau_2 = 1$), or total concentration of the sorption site available $\bar{S}$ (when $\tau_3 = 1$). It is worth mentioning that only one kind of sorption is assumed to exist in porous media in this work. Data mismatch can also be considered in MSE form, as follows:

$$MSE_{data1} = \frac{1}{N_{obs1}}\sum_{n=1}^{N_{obs1}}(\hat{h}_n - h_n)^2 \quad (12)$$

$$MSE_{data2} = \frac{1}{N_{obs2}}\sum_{n=1}^{N_{obs2}}(\hat{C}_n - C_n)^2 \quad (13)$$

where $N_{obs1}$ and $N_{obs2}$ denote the number of observation data of labeled data $h$ and $C$, respectively. In this work, the labeled data are obtained from numerical simulation.

For clarity, the loss function of TgU-net for hydraulic head and contaminant can be written as **Eq. (14)** and **Eq. (15)**, respectively:

$$L_h = MSE_{data1} + \beta_1 MSE_{pde1} \quad (14)$$
$$L_C = MSE_{data2} + \beta_2 MSE_{pde2} \quad (15)$$

where $\beta_1$ and $\beta_2$ are hyper-parameters to control the weight of the residual of PDEs. In this work, $\beta_1$ and $\beta_2$ are both set to be 100. TgU-net-C will predict contaminant fields for all time-steps at once. In addition, the training process of TgU-net-h and TgU-net-C are the same, as shown in **Fig. 2**.

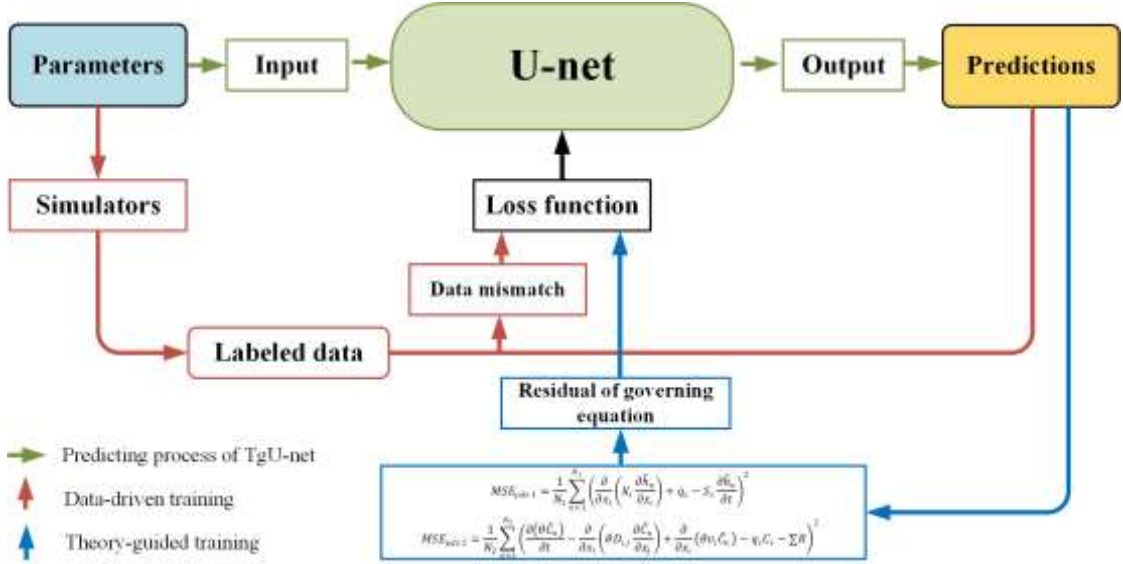

**Figure 2.** Training process of TgU-net.

**2.3 Iterative ensemble smoother**

In this work, the iterative ensemble smoother (IES) is employed for inverse modeling of the contaminant transport problem. IES is a kind of ensemble-based data assimilation method. Traditional gradient descent methods rely on gradient information, which is hard to obtain for subsurface problems. In the IES method, however, the gradient is replaced by the covariance, which is easier to implement. The IES method has been applied broadly in inverse modeling of petroleum engineering and hydraulic problems (Oliver et al., 2008; Chang et al., 2017; Chang and Zhang, 2019; He et al., 2021; Wang et al., 2021b). A brief introduction is given below.

A physical model can be presented as follows (Oliver et al., 2008):

$$\boldsymbol{d} = g(\boldsymbol{m}) \qquad (16)$$

where $g(\cdot)$ is the physical system; $\boldsymbol{m}$ is the parameters of $g(\cdot)$; and $\boldsymbol{d}$ denotes the response.

If the parameters $\boldsymbol{m}$ is unknown, the process of estimating $\boldsymbol{m}$ can be regarded as an inverse problem. One way to solve this inverse problem is to maximize the conditional probability density function (PDF) p($\boldsymbol{m}|\boldsymbol{d}^{obs}$) (Chang et al., 2017). $\boldsymbol{d}^{obs}$ denotes the observation data. Following Bayes' rule, it is equivalent to minimizing **Eq. (17)** if the prior PDF of model parameters, model error, and measurement error all follow a Gaussian distribution (Oliver et al., 2008):

$$O(\boldsymbol{m}) = \frac{1}{2}\left(\boldsymbol{d}^{obs} - g(\boldsymbol{m})\right)^T C_D^{-1}\left(\boldsymbol{d}^{obs} - g(\boldsymbol{m})\right) \\ + \frac{1}{2}(\boldsymbol{m} - \boldsymbol{m}^{pr})^T C_M^{-1}(\boldsymbol{m} - \boldsymbol{m}^{pr}) \qquad (17)$$

where $C_D$ denotes the measurement error covariance matrix; $\boldsymbol{m}^{pr}$ denotes the prior parameters; and $C_M$ denotes the covariance matrix of $\boldsymbol{m}^{pr}$. **Eq. (17)** can be optimized through an iterative Gauss-Newton formula, as shown in **Eq. (18)** (Chen and Oliver, 2013):

$$m_{l+1} = m_l - [(1+\lambda_l)C_M^{-1} + G_l^T C_D^{-1} G_l]^{-1}$$
$$[C_M^{-1}(m_l - m^{pr}) + G_l^T C_D^{-1}(g(m_l) - d^{obs})] \quad (18)$$

where $l$ denotes the iterative index; $\lambda_l$ denotes the hyper-parameter to mitigate the influence of large data mismatch in early iterations; and $G_l$ denotes the sensitivity matrix of $m_l$ at iterative step $l$. In the ensemble methods, a group of realizations of model parameters that satisfy the Gaussian distribution is generated for representing the uncertainty. Moreover, Chang et al. (2017) proposed to approximate the sensitivity matrix with the covariance matrix to simplify the calculation process. The employed IES method can be written as follows:

$$m_{l+1,j} = m_{l,j} - \frac{1}{1+\lambda_l}\left[C_{M_l} - C_{M_l D_l}\left((1+\lambda_l)C_D + C_{D_l D_l}\right)^{-1} C_{D_l M_l}\right] C_M^{-1}(m_{l,j} - m_j^{pr})$$
$$-C_{M_l D_l}\left((1+\lambda_l)C_D + C_{D_l D_l}\right)^{-1}\left(g(m_{l,j}) - d_j^{obs}\right), \quad j = 1, \ldots, N_e \quad (19)$$

where $j$ is the index of realization; $N_e$ is the total number of realizations; $m_{l,j}$ denotes the estimated parameter of the $j$th realization at iterative step $l$; $C_{M_l}$ is the covariance matrix of $l$th model parameters; $C_{D_l,D_l}$ is the $l$th covariance matrix of model response; and $C_{D_l,M_l}$ is the $l$th covariance matrix of model response and parameters.

For the inverse modeling process, the forward model $g(\cdot)$ is usually implemented by numerical simulation. Since a large number of model executions is usually required in the IES method, traditional numerical simulation-based implementation may be computationally prohibitive. Herein, the proposed TgU-net is employed as a surrogate to approximate the forward model $g(\cdot)$ to reduce computational cost. The implementation process of the proposed algorithm is shown in **Fig. 3**, where $\varepsilon$ denotes the minimum RMSE threshold for $d_i$ from $g(m_i)$ and $d^{obs}$. For TgU-net-h, $m_i$ denotes hydraulic conductivity fields. For TgU-net-C, $m_i$ denotes information of contaminant fields, such as velocity fields obtained from $h$, sorption type indicators $\tau_1$, $\tau_2$, and $\tau_3$, and source concentration. The $m_i$ should be initialized prior to the first iteration. The updated model parameters may convergence to the real value with iteration.

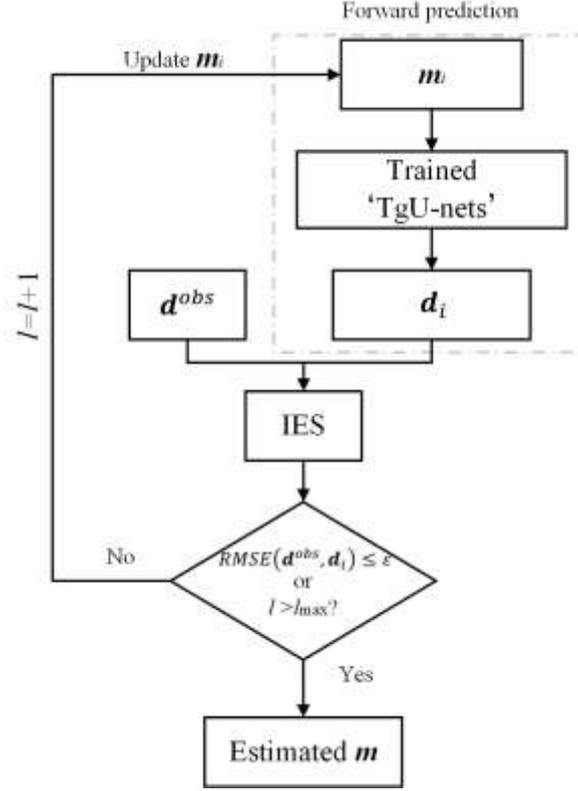

**Figure 3.** Implementation process of TgU-net and IES method.

## 3. Case Study

In this section, before discussing the inverse modeling results, the performance of TgU-net surrogate models is first tested. Two surrogate models are respectively constructed for $h$ and $C$ fields. Parameters of the contaminant source and hydraulic conductivity fields should be included in the input of the corresponding TgU-net. The input of TgU-net-h is a 3D hydraulic conductivity field. For TgU-net-C, the input should consist of seepage $v_i$, location and strength of contaminant source, $\tau_i$, $\gamma_1$, and $\gamma_2$. The $v_i$ is calculated through finite difference methods according to **Eq. (3)**, where $h$ can be obtained from TgU-net-h or a simulator. The sequential structure of the mentioned two TgU-nets is shown in **Fig. 4**. Information of the contaminant transport problem is integrated into different channels of a 3D block with the same size as the contaminant field. Reference values can be obtained from simulators and further used to calculate data mismatch. For TgU-net-h, the prediction is a 3D hydraulic head field. TgU-net-C should predict 3D contaminant fields of all time-steps to calculate the time gradient term that is required for incorporating the PDE constraint. In the training process, the residual of the governing equation and data mismatch should decrease simultaneously.

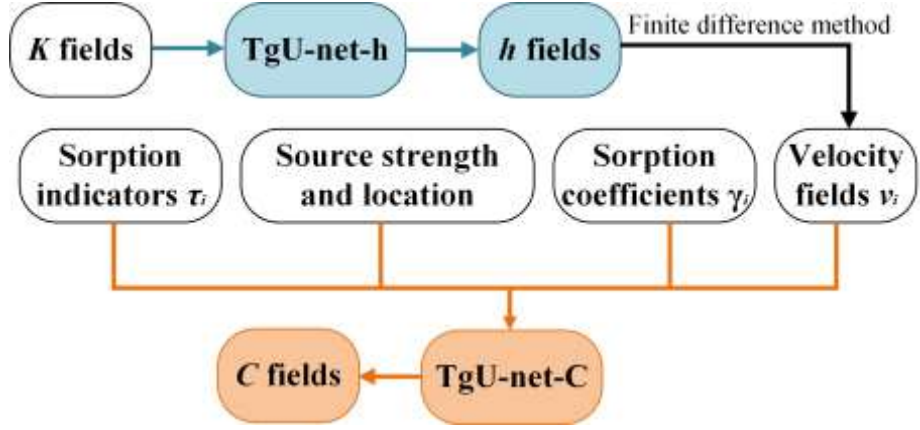

**Figure 4.** Sequential structure of TgU-net-h and TgU-net-C.

In this work, $K$ fields are generated through the Karhunen–Loeve expansion (KLE) method (Zhang and Lu, 2004), as shown in **Eq. (20)**:

$$lnK(\mathbf{x}) \approx \overline{lnK(\mathbf{x})} + \sum_{i=1}^{n_{tr}} \sqrt{\omega_i} f_i(\mathbf{x}) \xi_i(\eta) \tag{20}$$

where $\overline{lnK(\mathbf{x})}$ denotes the mean of $lnK(\mathbf{x})$; $n_{tr}$ denotes the number of truncated terms; $\omega_i$ denotes eigenvalues; $f_i(\mathbf{x})$ denotes eigenfunctions; $\eta$ denotes variables from probability spaces; and $\xi_i(\eta)$ denotes orthogonal Gaussian variables.

The studied domain is a 248 [L] × 488 [L] × 120 [L] block. It is discretized uniformly into 31 × 61 × 15 grids with $\Delta x = \Delta y = \Delta z = 8$ [L]. For the flow field, the left side of the studied domain is the entrance and the right one is the exit, respectively, taking hydraulic head values of 50 [L] and 0 [L], respectively. The lower and upper boundaries are both set to be impervious. Boundary conditions of **Eq. (1)** and some parameters of **Eq. (2)** are fixed. Values of fixed parameters are displayed in **Table 1**. The injected contaminant source is located in a 3 × 3 matrix that runs through from the upper to the lower layer. The $h$ is assumed to be a steady state flow field, and thus the term $S_s \frac{\partial \hat{h}_n}{\partial t}$ in **Eq. (9)** can be ignored. The duration of dynamic $C$ fields is 360 [T], and the time-step is set to be 6 [T]. Herein, reference $h$ and $C$ are obtained from software MODFLOW and MT3DMS, respectively (Zheng and Wang, 1999).

**Table 1.** Values of fixed parameters of contaminant fields.

| Parameter | $\theta$ | $\rho_d$ | $\alpha_L$ | $\alpha_{TV}$ | $\alpha_{TH}$ |
|---|---|---|---|---|---|
| Value | 0.3 | 1.7 | 80 | 8 | 8 |

The TgU-net surrogate models are constructed and run in the Pytorch environment, a python-based deep-learning framework (Imambi et al., 2021). In order to avoid vanishing gradient, the LeakyReLU function is chosen as the activation function in each layer of TgU-net (Lakshmi et al., 2021). The learning rate is set to be 1e-4. The total training epoch is 2,000. If the prediction error fluctuates within an acceptable range, training will be terminated in advance. Weights and bias of TgU-net are initialized via the Xavier uniform method. AdamW (Loshchilov and Hutter, 2017) is chosen as the optimizer. It is an improvement over the traditional Adam (Kingma and Ba, 2015), and it has a faster convergence rate and superior stability. All programs are processed on a NVIDIA GeForce RTX 2080 Ti GPU. Criteria, including root-mean-square error (RMSE) and $R^2$ score, are employed to quantitatively evaluate the performance of the trained TgU-net, which are expressed as follows:

$$RMSE = \sqrt{\frac{1}{N}\sum_{i=1}^{N}(\hat{u}_i - u_i)^2} \quad (21)$$

$$R^2 = 1 - \frac{\sum_{i=1}^{N}(\hat{u}_i - u_i)^2}{\sum_{i=1}^{N}(u_i - \overline{u_i})^2} \quad (22)$$

where $\hat{u}_i$ denotes predictions; $u_i$ denotes reference data; and $N$ is the number of reference data.

*K* fields out of training configurations are fed to the trained surrogate models to validate accuracy and generalizability in section 3.1. In section 3.2, observations of both *h* and *C* fields are utilized to identify physical processes and unknown characteristics. In addition, some sparsity-promoting techniques are utilized and discussed. Finally, the influence of strong nonlinearity and information volume on data assimilation results is explored in section 3.3. Specifically, only observation data from the *C* field are analyzed in the IES method.

**3.1 Accuracy, extrapolability, and generalizability of TgU-net surrogate models**

In this section, fitting ability, generalizability, and extrapolability of the constructed TgU-net-based surrogate models are tested with sparse data points. Moreover, the corresponding traditional U-nets are trained in the same configurations as comparisons. For TgU-net-h, the correlation length of *lnK* along each side is 0.2 times the corresponding geometrical length. In the training stage, $\overline{lnK(\mathbf{x})}$ is set to be 1, and the variance of *lnK* is 0.5. The *h* field is considered as a steady state flow field. 1,000 labeled and 1,000 virtual (no observation points) realizations are utilized to train TgU-net-h and U-net-h. The labeled realizations are obtained from running the simulator. Furthermore, for label realizations, observation data are taken at every other point on the horizontal plane, running through the uppermost and lowermost layers, amounting to 6,750 (15 × 30 × 15) data points. The virtual realizations are used for enforcing PDE constraints, and no numerical simulations are required. For the virtual realizations, all points of the whole domain are used as collocation points to calculate **Eq. (9)** (i.e., 27 × 57 × 11, amounting to 16,929 collocation points). The observation data are usually sparse in practice, while the collocation points can be virtually intact. In addition, the traditional U-net is trained only in a data-driven scheme. Training TgU-net-h and U-net-h 2,000 epochs takes 26,580 s and 14,074 s, respectively.

After training, the test *K* field has the same setup as the training ones, but is out of training range, as shown in **Fig. 5**, and is inputted to test the fitting ability. Predicted *h* fields of U-net-h and TgU-net-h are presented in **Fig. A1** and **Fig. A2**, respectively, in Appendix A. It can be seen that TgU-net outperforms U-net. $R^2$ score and RMSE of 15 layers are shown in **Fig. 6**. It is obvious that the trained traditional U-net cannot predict the whole domain satisfactorily due to the sparse observation data. For the locations without collected data, the prediction of U-net is poor, indicating that the traditional U-net could not learn the inherent governing equation. With the assistance of theory-guidance, the prediction of TgU-net is much more continuous, leading to a significant increase in accuracy. In order to test the effectiveness of data points, RMSEs of these points predicted by TgU-net-h and U-net-h are calculated, as shown in **Fig. 7**, and the corresponding values are presented in **Table A1** in Appendix A. Although the traditional U-net may achieve satisfactory predictions with abundant observation data, data are usually hard to obtain for subsurface problems. The advantage of TgU-net lies in its stable prediction ability even in the case of sparse data. The prediction accuracy of TgU-net can be further improved due to the incorporation of governing equation constraints. Moreover, generalizability and extrapolability of the constructed TgU-net are evaluated through unseen configurations. For instance, the mean of *lnK* increased from 1 to 1.5 while the variance remained 0.5, or *lnK* with mean value of 1 and variance of 0.8, as shown in **Fig. 8**. $R^2$ score and RMSE of predictions from TgU-net-h and U-net-h are presented in **Fig. 9**. It can be seen that TgU-net can still obtain satisfactory and stable predictions when dealing with unseen configurations. TgU-net can effectively achieve results by obeying the governing equation; whereas, this is impossible for the data-driven U-net. It is worth noting that the prediction performance of the data-driven U-net is similar under these three *K* fields, and the curves of RMSE are in the shape of an "M". To explain this, we calculated the variance of the *lnK* participating in the prediction, as shown in **Fig. 10**. The results reveal that the data-driven U-net is sensitive to the variance of *lnK*. In addition, especially when the variance of *lnK* is larger, the results become markedly worse, indicating that the data-driven U-net cannot capture large variability of *lnK*. In contrast, due to the effect of theory-guidance, the TgU-net can accommodate *lnK* fields with different variance. The findings above show that the TgU-net-based surrogate model for predicting hydraulic heads possesses satisfactory fitting ability, generalizability, and extraplolability. It can also be further utilized to solve inverse problems with data assimilation algorithms.

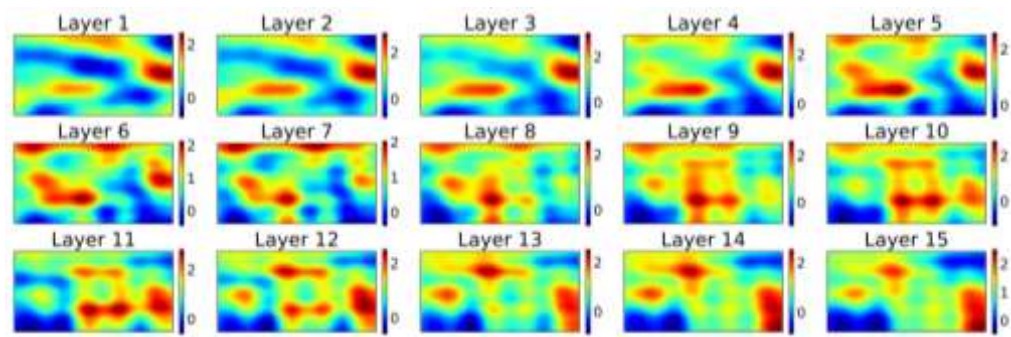

**Figure 5.** Demonstration of the test *lnK* field.

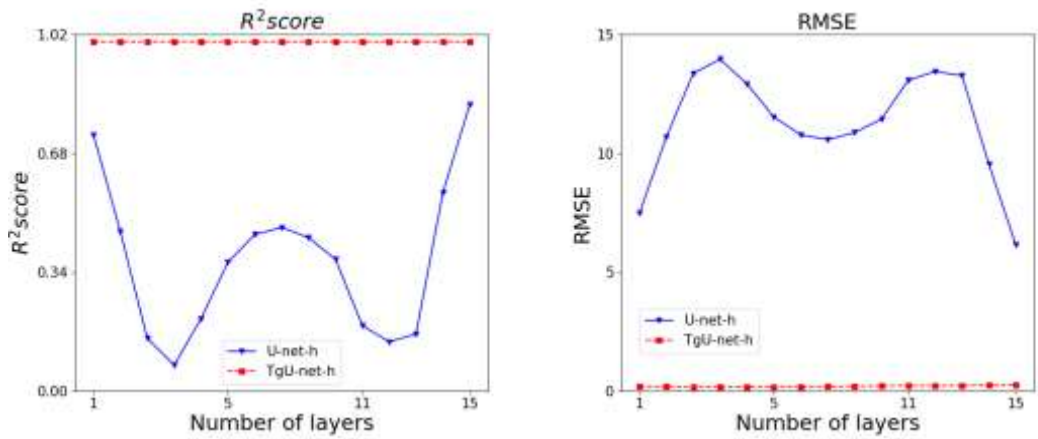

**Figure 6.** Comparison of $R^2$ score and RMSE of 15 layers.

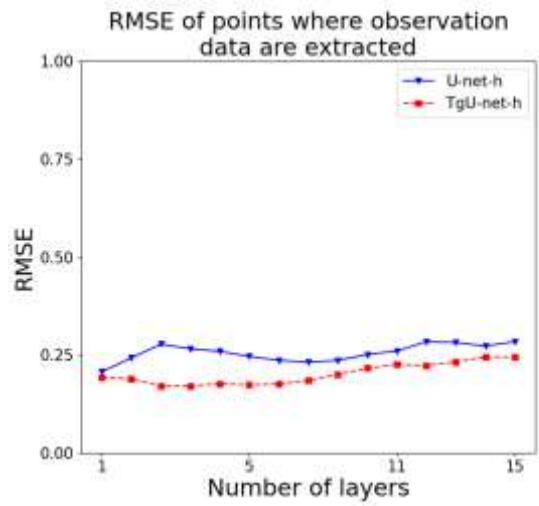

**Figure 7.** Comparison of RMSE of data points.

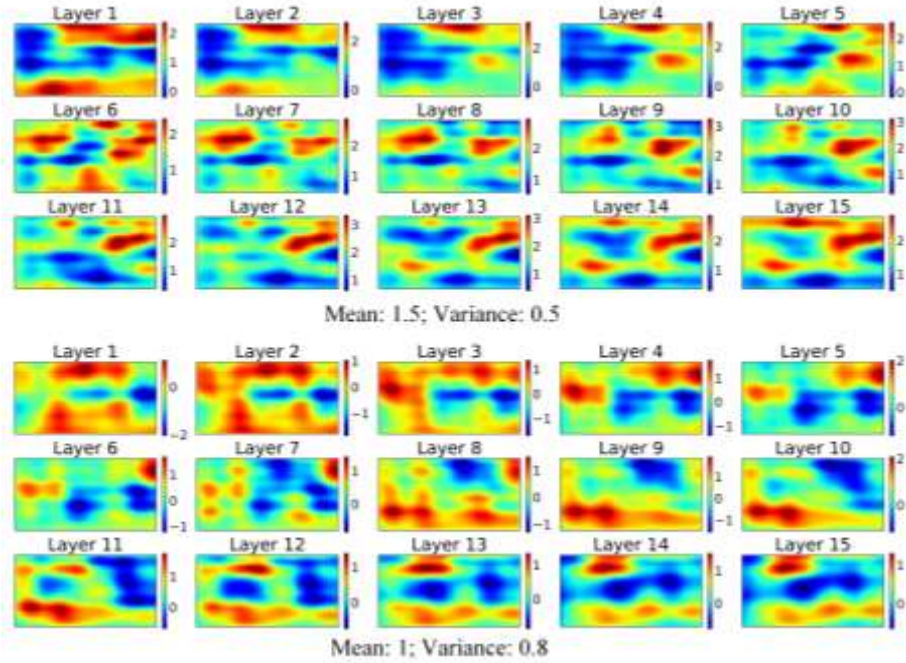

**Figure 8.** Demonstration of test *lnK* fields for evaluating generalization and extrapolation capacities of TgU-net and U-net.

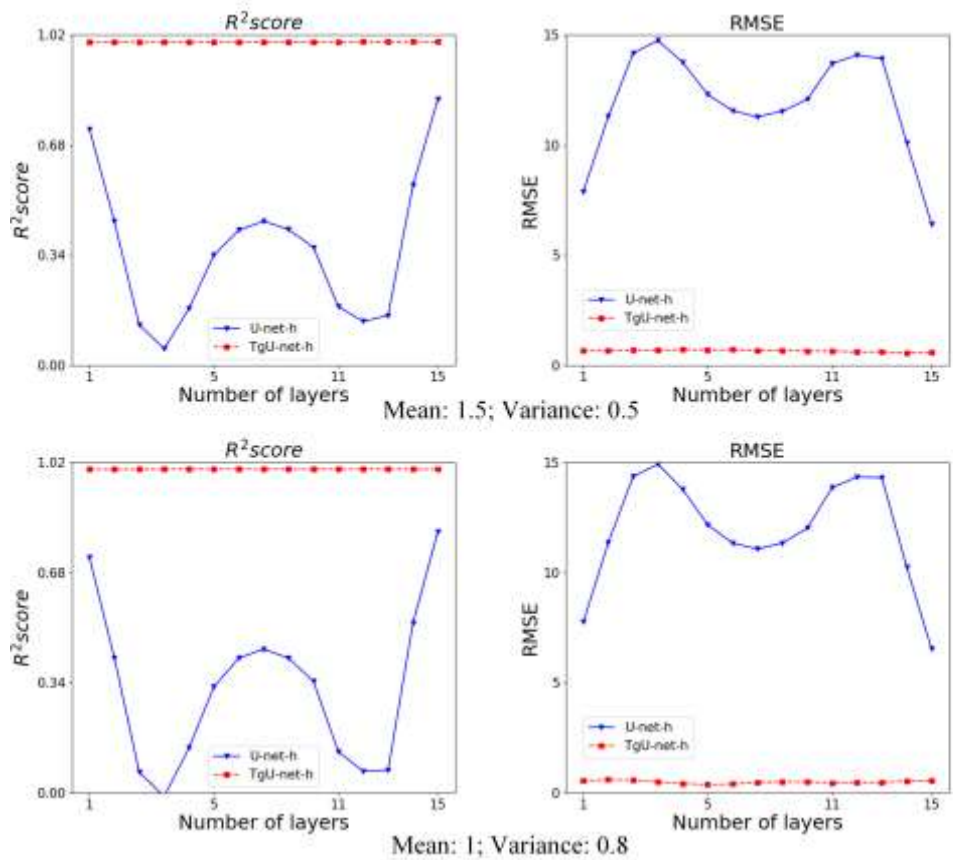

**Figure 9.** $R^2$ score and RMSE of predictions under *lnK* fields with out of training range configurations.

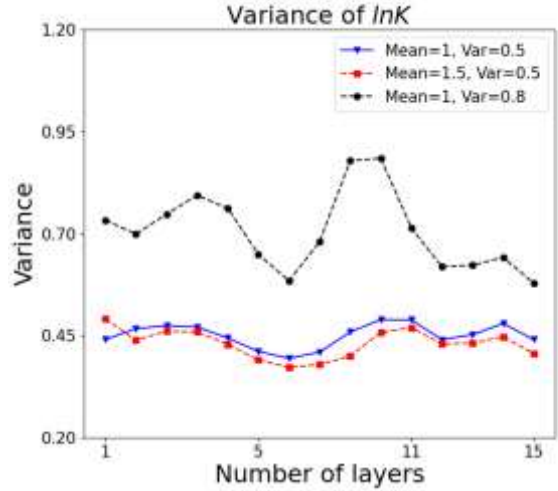

**Figure 10.** Variance of each layer of hydraulic conductivity fields.

The performances of TgU-net and traditional U-net for contaminant fields are also assessed. The input of TgU-net-C consists of seepage $v_x$, $v_y$, $v_z$, location and strength of contaminant source, $\tau_i$, $\gamma_1$, and $\gamma_2$. Prior distributions of contaminant source parameters are shown in **Table 2.** The method of sampling data points and collocation points at every other time-step is the same as that of TgU-net-h, i.e., 202,500 data points (6,750 × 30 time-steps) and 507,870 collocation points (16,929 × 30 time-steps). 1,000 virtual and 1,000 labeled training realizations result in approximately 32,280 s of training time. The training time of the conventional U-net-C is 14,761 s. Since, in this work, sorption is considered to be a potential process with three possible types, the training realizations contain different numbers of realizations for the cases of no sorption, linear, Freundlich, and Langmuir sorption, respectively. The training numbers of no sorption, linear, Freundlich, and Langmuir sorption realizations, RMSE mean, and variance of 100 predictions are presented in **Table 3**. The training data are nearly uniformly distributed. The prediction accuracy of TgU-net-C decreases slightly with the complexity of sorption type, but is within the acceptable range. If there is no sorption, the dynamic characteristic of contaminant transport is stronger and the data change is more drastic, which makes it more difficult to be learned than that when the retardation factor is present. Furthermore, the retardation effect of Langmuir sorption is the weakest, resulting in a similar accuracy to that of no sorption. **Fig. 11** displays concentration curves at time-step 60 of these four situations to further demonstrate this phenomenon. The results of the traditional U-net presented in **Table 3** are not satisfactory, indicating that the data-driven neural network with sparse data cannot learn the inherent governing equation. The variability of training data has a strong influence on the prediction results of U-net-C. Therefore, we can see that incorporating governing equation constraints in the training process can improve the prediction accuracy and generalization ability of NNs when faced with data with strong variability.

Table 2. Prior distributions of contaminant source parameters.

| Parameters | $x$ | $y$ | $q_s C_s$ | $K_d$ |
|---|---|---|---|---|
| Prior | [80,160] | [104,460] | [10,20] | [0.1,0.6] |
| Parameters | $K_f$ | $K_l$ | $a$ | $\bar{S}$ |
| Prior | [0.1,0.6] | [30,200] | [0.3,1.3] | [0.003,0.015] |

**Note:** $x$ and $y$ denote the x- and y-coordinate of the 3 × 3 matrix center point, respectively.

Table 3. Number of training realizations and predicted RMSE.

| | | No sorption | Linear | Freundlich | Langmuir |
|---|---|---|---|---|---|
| Training realization | Labeled | 228 | 258 | 255 | 259 |
| | Virtual | 265 | 237 | 245 | 253 |
| RMSE (U-net) | Mean | 2.6284 | 1.6447 | 2.1678 | 2.3753 |
| | Variance | 1.1099 | 0.4094 | 0.7687 | 0.9814 |
| RMSE (TgU-net) | Mean | 1.106e-1 | 8.095e-2 | 9.278e-2 | 9.973e-2 |
| | Variance | 5.699e-3 | 4.551e-3 | 3.965e-3 | 4.473e-3 |

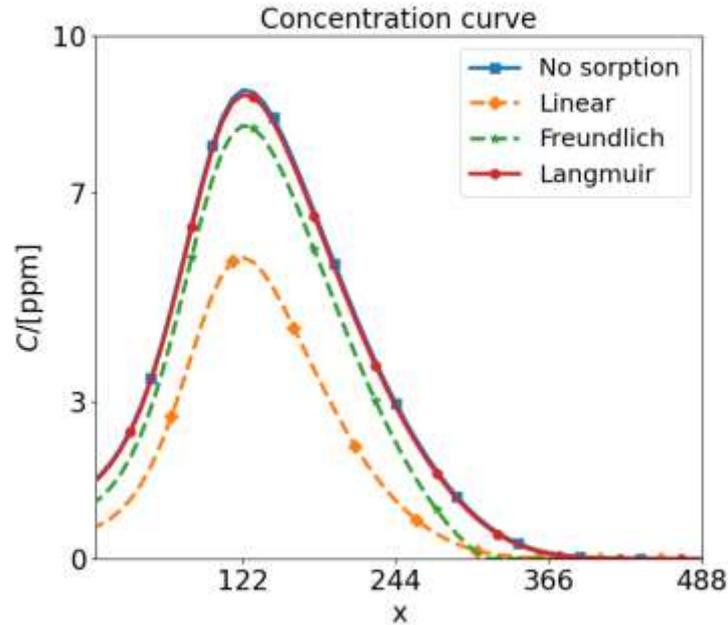

**Figure 11.** Concentration curve of four sorption types (the first layer, y = 150, time-step 60).

Seepages $v_x$, $v_y$, and $v_z$ obtained from ln$K$ fields out of training range (mean: 1, variance: 0.8; mean: 1.5, variance: 0.5) are used to validate the generalization and extrapolation ability of TgU-net-C. Histograms of $R^2$ score and RMSE of each predicted layer are shown in **Fig. 12**. The mean and variance of $R^2$ score and RMSE are displayed in **Table 4**. 2,000 points at time-step 20, 40, and 60 are randomly selected to further demonstrate the fitting ability of U-net-C and TgU-net-C, as shown in **Fig.**

**13**. It can be seen that the $R^2$ score of TgU-net is close to 1 and the RMSE is concentrated around 0, demonstrating satisfactory performance of TgU-net. Global predictions of TgU-net can not only honor the observation data, but also obey underlying governing equations due to the theory-guidance training scheme. However, almost all extracted points in **Fig. 13 (a)** are not on the line y = x, indicating that the purely data-driven U-net by sparse observation is unreliable in terms of extrapolation. Moreover, forward prediction computation time of MODFLOW, MT3DMS, TgU-net-h, and TgU-net-C is recorded, as shown in **Fig. 14**. The advantage of computation efficiency of TgU-net-based surrogate models is more evident when a substantial number of model executions is required. In summary, TgU-net-based surrogate models possess more satisfactory generalization and extrapolation abilities compared to the traditional data-driven U-net. TgU-net surrogates have the potential to be forward prediction models in data assimilation methods due to their accuracy and efficiency.

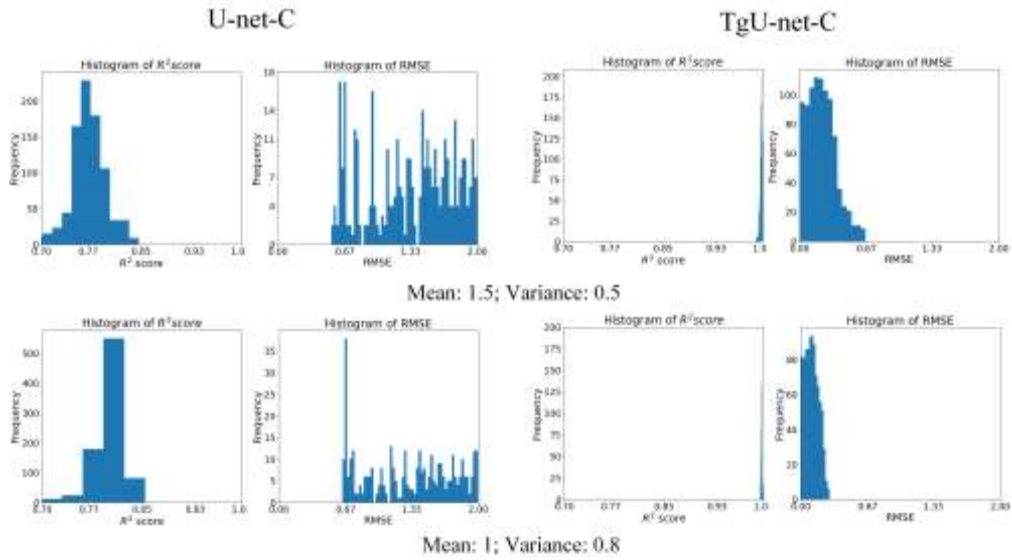

**Figure 12.** Comparison of $R^2$ score and RMSE histograms.

**Table 4**. Comparison of $R^2$ score and RMSE of traditional U-net-C and TgU-net-C.

| Condition of ln$K$ | U-net-C | | | |
|---|---|---|---|---|
| | $R^2$ score | | RMSE | |
| | Mean | Variance | Mean | Variance |
| Mean: 1.5 Variance: 0.5 | 0.7135 | 0.2819 | 2.1810 | 0.7817 |
| Mean: 1 Variance: 0.8 | 0.7353 | 4.6900e-2 | 2.1292 | 0.6910 |
| | TgU-net-C | | | |
| | $R^2$ score | | RMSE | |
| | Mean | Variance | Mean | Variance |
| Mean: 1.5 Variance: 0.5 | 0.9976 | 2.8267e-6 | 0.2248 | 1.9232e-2 |
| Mean: 1 Variance: 0.8 | 0.9984 | 1.1453e-7 | 0.1153 | 4.5188e-3 |

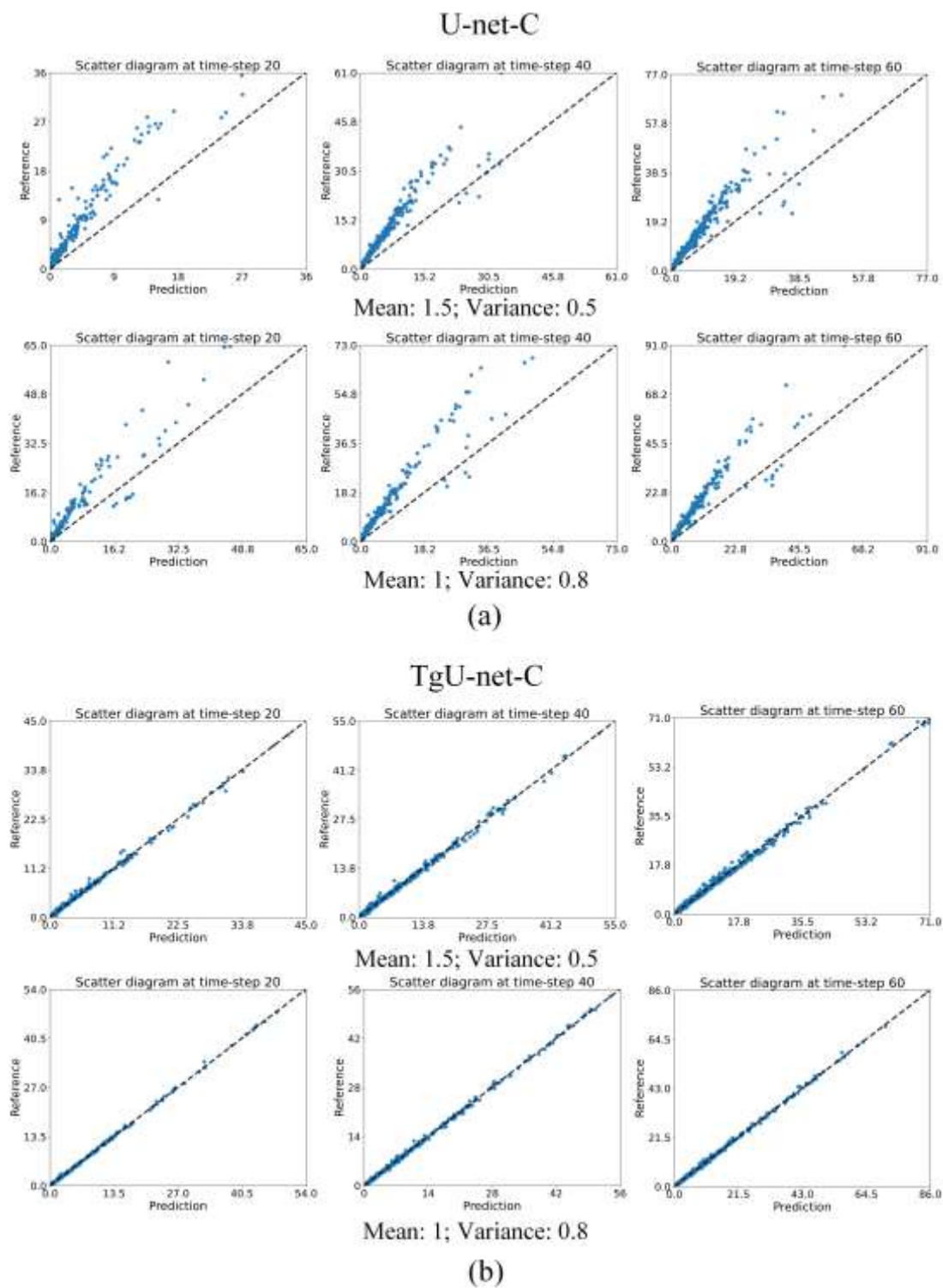

**Figure 13.** Scatter diagrams of U-net-C (a) and TgU-net-C (b) predictions.

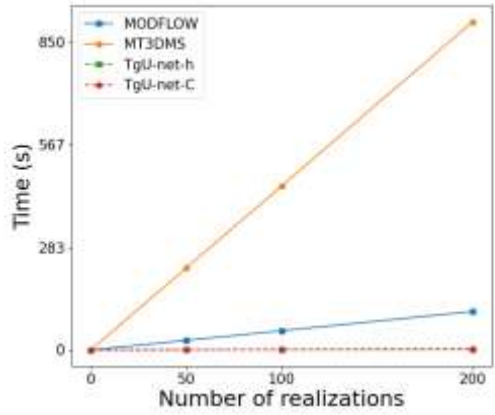

**Figure 14.** Comparison of computation time.

**3.2 Identification of physical processes and unknown characteristics**

In this section, the trained TgU-net-based surrogate models are combined with the IES method for inverse modeling. Firstly, by combining TgU-net-h and IES, observation data of hydraulic head fields are used to estimate $K$ fields. Then, the velocity fields with respect to the updated $K$ fields are inputted to TgU-net-C. Finally, by combining TgU-net-C and IES, observation data of contaminant concentration are used to estimate unknown source parameters and physical processes. Sampling positions of observation data in each layer are shown in **Fig. 15**, whether in the hydraulic head or the contaminant concentration fields. For performing IES, the number of iteration times is 15, and the ensemble size is set to be 200.

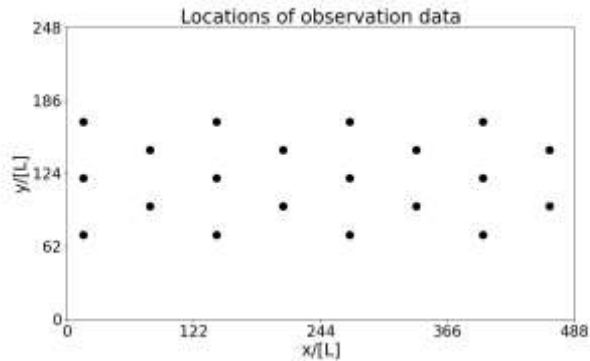

**Figure 15.** Sampling locations of observation data in each layer.

Firstly, hydraulic conductivity fields are inversed with observation $h$ data without noise. The reference ln$K$ field is shown in **Fig. 16**. RMSEs of the mean of initialized and estimated ln$K$ in the IES method are presented in **Table 5**. It can be seen that the RMSEs of the estimated ln$K$ decrease largely, indicating that it is more similar to the reference field. In order to test the effectiveness of the estimated ln$K$ field, the responding $h$ field is calculated by the trained TgU-net-h, and presented as **Fig. B1** in Appendix B. The $R^2$ score and RMSE of the predicted $h$ field are shown in **Table 6**.

The $R^2$ score is close to 1 and the RMSE is close to 0, indicating the accuracy of the estimated *K* field, and further reflecting the reliability of the IES method and the TgU-net surrogate model.

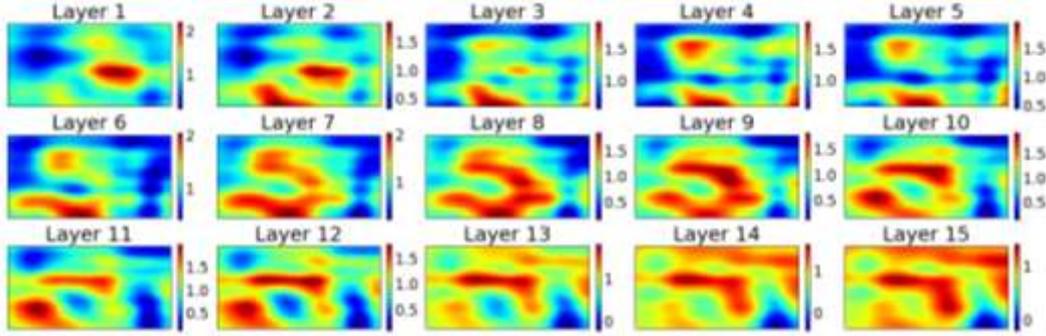

**Figure 16.** Demonstration of the reference ln*K*.

**Table 5**. RMSE of the mean of initialized and estimated ln*K* in each layer.

| | Layer 1 | Layer 2 | Layer 3 | Layer 4 | Layer 5 |
|---|---|---|---|---|---|
| | 0.3374 | 0.2774 | 0.2692 | 0.3149 | 0.3113 |
| Initialized | Layer 6 | Layer 7 | Layer 8 | Layer 9 | Layer 10 |
| ln*K* | 0.3286 | 0.3530 | 0.3378 | 0.3202 | 0.3392 |
| | Layer 11 | Layer 12 | Layer 13 | Layer 14 | Layer 15 |
| | 0.3607 | 0.3286 | 0.3169 | 0.3199 | 0.2798 |
| | Layer 1 | Layer 2 | Layer 3 | Layer 4 | Layer 5 |
| | 0.0511 | 0.0429 | 0.0343 | 0.0366 | 0.0434 |
| Estimated | Layer 6 | Layer 7 | Layer 8 | Layer 9 | Layer 10 |
| ln*K* | 0.0335 | 0.0343 | 0.0466 | 0.0468 | 0.0452 |
| | Layer 11 | Layer 12 | Layer 13 | Layer 14 | Layer 15 |
| | 0.0548 | 0.0491 | 0.0429 | 0.0457 | 0.0396 |

**Table 6**. $R^2$ score and RMSE of predicted *h* field through estimated ln*K* field.

| | Layer 1 | Layer 2 | Layer 3 | Layer 4 | Layer 5 |
|---|---|---|---|---|---|
| | 0.9992 | 0.9993 | 0.9991 | 0.9992 | 0.9991 |
| $R^2$ score | Layer 6 | Layer 7 | Layer 8 | Layer 9 | Layer 10 |
| | 0.9991 | 0.9990 | 0.9991 | 0.9992 | 0.9992 |
| | Layer 11 | Layer 12 | Layer 13 | Layer 14 | Layer 15 |
| | 0.9992 | 0.9993 | 0.9993 | 0.9994 | 0.9994 |
| | Layer 1 | Layer 2 | Layer 3 | Layer 4 | Layer 5 |
| | 0.3858 | 0.3747 | 0.4130 | 0.4088 | 0.4292 |
| RMSE | Layer 6 | Layer 7 | Layer 8 | Layer 9 | Layer 10 |
| | 0.4265 | 0.4366 | 0.4275 | 0.4209 | 0.4034 |
| | Layer 11 | Layer 12 | Layer 13 | Layer 14 | Layer 15 |
| | 0.3979 | 0.3765 | 0.3769 | 0.3445 | 0.3441 |

The influence of noise is discussed here. In general, observation data are usually accompanied by noise. Herein, we test the noise resistance ability of the TgU-net-based IES method. The way of adding noise is as follows (Chang and Zhang, 2019):

$$D(x,y,z,t) = D(x,y,z,t) \times (1 + \delta \times e) \tag{23}$$

where $D(x,y,z,t)$ denotes the observation data; $\delta$ denotes the level of added noise; and $e$ is the uniformly random variable with values in the interval [-1,1], and with the same data volume as $D(x,y,z,t)$. In this work, 1%, 5%, and 10% noise are considered to test the performance of the proposed method. It is worth mentioning that the initialized ln$K$ fields in the IES method for the cases of different noise levels are the same. The RMSE of estimated ln$K$ fields are presented in **Table 7**. With the increase of noise level, the RMSE of estimated ln$K$ fields obviously decrease. The existence of noise may reinforce uncertainty and increase the difficulty of solving inverse problems. However, the results of different noise levels are still acceptable, demonstrating that the proposed method possesses a satisfactory ability to resist noise. In order to further determine the validity of the estimations with different noise levels, these $K$ fields are inputted into the trained TgU-net-h to calculate the accuracy of the hydraulic head, as shown in **Fig. 17**. The predicted results can match the reference well, indicating the robustness of the proposed method.

**Table 7**. RMSE of the mean of estimated ln$K$s with different noise levels.

| | Layer 1 | Layer 2 | Layer 3 | Layer 4 | Layer 5 |
|---|---|---|---|---|---|
| | 0.0511 | 0.0429 | 0.0343 | 0.0366 | 0.0434 |
| Clean data | Layer 6 | Layer 7 | Layer 8 | Layer 9 | Layer 10 |
| | 0.0335 | 0.0343 | 0.0466 | 0.0468 | 0.0452 |
| | Layer 11 | Layer 12 | Layer 13 | Layer 14 | Layer 15 |
| | 0.0548 | 0.0491 | 0.0429 | 0.0457 | 0.0396 |
| | Layer 1 | Layer 2 | Layer 3 | Layer 4 | Layer 5 |
| | 0.0751 | 0.0638 | 0.0449 | 0.0497 | 0.0625 |
| 1% noise | Layer 6 | Layer 7 | Layer 8 | Layer 9 | Layer 10 |
| | 0.0501 | 0.0471 | 0.0619 | 0.0605 | 0.0578 |
| | Layer 11 | Layer 12 | Layer 13 | Layer 14 | Layer 15 |
| | 0.0687 | 0.0584 | 0.0532 | 0.0589 | 0.0529 |
| | Layer 1 | Layer 2 | Layer 3 | Layer 4 | Layer 5 |
| | 0.1842 | 0.1572 | 0.1039 | 0.1160 | 0.1507 |
| 5% noise | Layer 6 | Layer 7 | Layer 8 | Layer 9 | Layer 10 |
| | 0.1613 | 0.1367 | 0.1352 | 0.1288 | 0.1255 |
| | Layer 11 | Layer 12 | Layer 13 | Layer 14 | Layer 15 |
| | 0.1423 | 0.1583 | 0.1673 | 0.1500 | 0.1225 |
| | Layer 1 | Layer 2 | Layer 3 | Layer 4 | Layer 5 |
| | 0.2450 | 0.1882 | 0.1573 | 0.1653 | 0.1773 |
| 10% noise | Layer 6 | Layer 7 | Layer 8 | Layer 9 | Layer 10 |
| | 0.2112 | 0.1874 | 0.1890 | 0.1799 | 0.1728 |
| | Layer 11 | Layer 12 | Layer 13 | Layer 14 | Layer 15 |
| | 0.1554 | 0.1620 | 0.1830 | 0.1661 | 0.1405 |

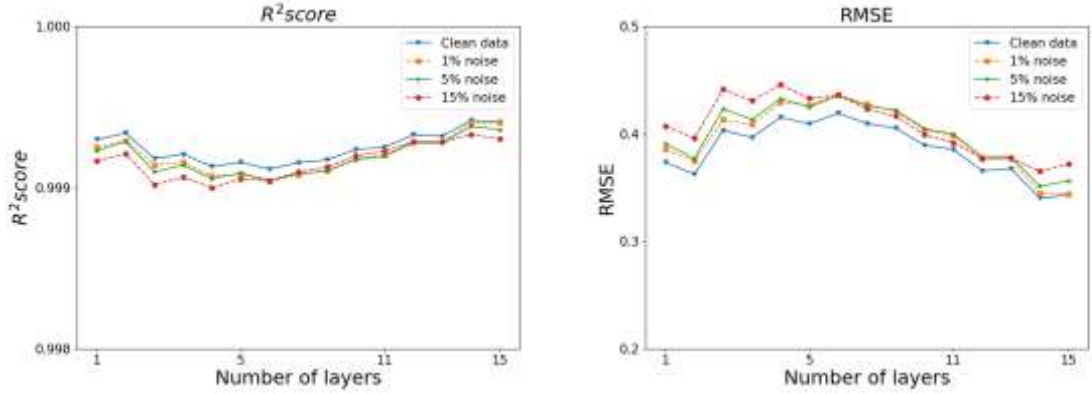

**Figure 17.** Curves of $R^2$ score and RMSE of predictions under *lnK* fields with different noise levels.

After the inversion of *K* fields with data from hydraulic head fields, the velocity fields $v_x$, $v_y$, and $v_z$ can be calculated and used as fixed input of the contaminant field surrogate model TgU-net-C. Then, the governing physical process and unknown source parameters can be inversed. The reference, initialized, and estimated source parameters of a contaminant convection-dispersion process with Langmuir sorption are summarized in **Table 8**. For convenience of statistics, $\gamma_1$ and $\gamma_2$ are regularized and limited between [0-1]. If the sorption type is identified to be Langmuir, then $\gamma_1$ is multiplied by 1,000 and $\gamma_2$ is divided by 100. Boxplots of initialized and estimated parameters are shown in **Fig. 18** to compare the results of convergence. The results of cases with linear and Freundlich sorption are displayed in **Fig. B2** and **B3**, respectively, in Appendix B. It is obvious that each contaminant parameter converges near the reference value and the variance is negligibly small, which indicates the reliability of the constructed surrogate models and the effectiveness of the IES method. Because the estimated value of $\tau_3$ is close to 1, Langmuir sorption is shown to be the proper empirical process of the investigated contaminant convection-dispersion problem. Results demonstrate that this proposed algorithm, i.e., combination of TgU-net surrogate models and the IES method, can be applied to identify the proper model for uncertain physical processes. Furthermore, the influence of noise on the results of inverting source parameters is also discussed. **Table 9** displays the convergence of estimated parameters under different noise levels of Langmuir sorption. The results of other sorption types are presented in **Table B1** and **B2** in Appendix B. With the increase of noise, the variance of estimated source parameters becomes larger and deviates from the reference value, and especially the indicators of sorption types. However, the error of parameters is still in the acceptable range, further indicating the satisfactory anti-noise ability of the proposed method. It is also worth noting that, although the prediction and extrapolation accuracy of linear sorption is higher than that of the other two kinds of sorption, the inversion results are worse. This may be because the observation data of Langmuir and Freundlich have large spatial and temporal variations, and thus more information can be expressed and the features of the contaminant field are more easily captured. In addition, smooth data may not reflect dynamic features well. If the observation data of contaminant fields with the mentioned sorption types are regularized simultaneously to [0-1], the change curve of linear sorption will approach a straight line, making inversion more challenging.

An interesting phenomenon can be observed, i.e., when the noise level exceeds 10%, the values of sorption indicators that are not supposed to exist are not negligibly small. For instance, in the case with Freundlich sorption, three inversed sorption indicators are 0.0203, 0.8921, and 0.1684, respectively. In fact, the influence of non-zero sorption terms on contaminant convection-dispersion cannot be ignored numerically, and the learned equation is biased. Herein, it is hypothesized that only one sorption type exists, and thus we can regard the sorption type with the maximum indicator value as the dominant empirical term. In order to further explore this problem, we count the retardation effect caused by the reference term (i.e., $\tau_1 = 0$, $\tau_2 = 1$, and $\tau_3 = 0$) and the estimated indicators in the presence of 10% noise (i.e., $\tau_1 = 0.0203$, $\tau_2 = 0.8921$, and $\tau_3 = 0.1684$) from 100 random points, as presented in **Fig. 19**. The histograms show that the estimated indicators $\tau_i$ have a similar retardation effect to the reference ones. The mean of the normally distributed reaction-term error is 9.7928e-4, and the variance is 5.4843e-9, proving that predictions of TgU-net-based surrogate models indeed fit the governing equations. Indeed, the underlying equations are adequately captured by TgU-net models. In the iterative process of the IES method, predictions of the estimated comprehensive equations (i.e., **Eq. (10)** and **(11)**) can fit the observation data well, and the existence of noise makes the problem more uncertain and non-unique. This result further illustrates the effectiveness of the proposed TgU-net algorithms.

In order to reduce the influence of noise on the uncertainty of equation selection, the sorption type indicators $\tau_i$ are processed in the form of impulse signals. By using high-pass filtering, low-frequency signals are filtered, and only the highest signal value is retained. The filtering threshold is set to be 0.5. Moreover, the sequential thresholded least-squares algorithm (LASSO) method is utilized to improve indicator sparsity. The core algorithm of the LASSO algorithm is that of incorporating the L1 norm constraint of the coefficient, as shown in **Eq. (23)** (Brunton et al., 2016; Schaeffer et al., 2017). $\alpha$ is respectively set to be 0.05 and 0.1. The results displayed in **Table B3** in Appendix B show that sparsity-promoting techniques can assist to remove unnecessary indicators, further reducing the uncertainty of the model and improving the convergence of unknown characteristics. A larger $\alpha$ in the LASSO method can help to obtain more sparse results. Moreover, the influence of noise on other parameters is significantly eliminated due to the decrease of model uncertainty.

$$Q(\boldsymbol{m}) = (y - X\boldsymbol{m})^2 + \alpha\|\boldsymbol{m}\|_1 \tag{23}$$

where $Q(\boldsymbol{m})$ denotes the target function of the LASSO method; $y$ and $X$ are the observation data and independent valuable, respectively; $\alpha$ denotes the penalty coefficient, controlling sparsity degree; and $\boldsymbol{m}$ denotes the parameters to be dealt with, herein, referring to the sorption type indicators.

Table 8. Reference, initialized, and estimated source parameters (Langmuir sorption).

| Parameters | Reference | Initialized | | Estimated | |
|---|---|---|---|---|---|
| | | Mean | Variance | Mean | Variance |
| $x$ | 12 | 12.9795 | 0.6370 | 12.0456 | 5.832e-4 |
| $y$ | 16 | 16.0104 | 0.7619 | 16.0501 | 4.375e-4 |
| $q_s C_s$ | 11.0516 | 11.0099 | 0.3099 | 11.0769 | 4.003e-4 |
| $\tau_1$ | 0 | 0.2700 | 0.1971 | 3.916e-3 | 1.182e-4 |
| $\tau_2$ | 0 | 0.2600 | 0.1924 | 1.137e-3 | 1.554e-4 |
| $\tau_3$ | 1 | 0.2100 | 0.1659 | 0.9948 | 3.023e-4 |
| $\gamma_1$ | 0.1742 | 0.1301 | 6.092e-3 | 0.1703 | 1.538e-5 |
| $\gamma_2$ | 0.6917 | 0.3287 | 0.1246 | 0.7028 | 2.482e-4 |

**Note:** $x$ and $y$ denote the coordinates of the center point of the contaminant source matrix; $q_s C_s$ is the source strength; $\gamma_1$ is the distribution coefficient $K_d$ (when $\tau_1 = 1$), or the Freundlich equilibrium constant $K_f$ (when $\tau_2 = 1$), or the Langmuir equilibrium constant $K_l$ (when $\tau_3 = 1$); and $\gamma_2$ denotes the Freundlich exponent $a$ (when $\tau_2 = 1$), or total concentration of the sorption site available $\bar{S}$ (when $\tau_3 = 1$). For the convenience of statistics, $\gamma_1$ and $\gamma_2$ are regularized and limited between [0-1]. If the sorption type is judged to be Langmuir, then $\gamma_1$ is multiplied by 1,000 and $\gamma_2$ is divided by 100.

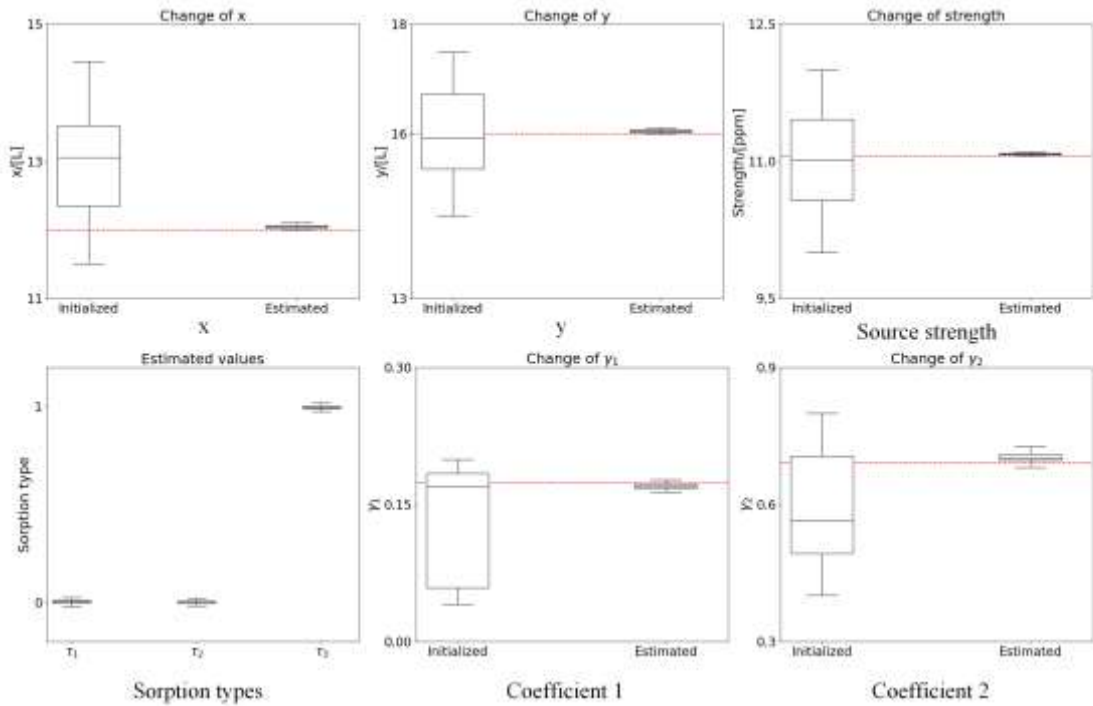

Figure 18. Comparison of convergence of initialized and estimated parameters (Langmuir sorption).

**Table 9.** Convergence of parameters under different noise levels (Langmuir sorption).

| Parameters | 1% noise | | 5% noise | | 10% noise | |
|---|---|---|---|---|---|---|
| | Mean | Variance | Mean | Variance | Mean | Variance |
| $x$ | 12.0461 | 5.579e-4 | 12.0479 | 5.387e-4 | 12.0529 | 6.367e-4 |
| $y$ | 16.0457 | 4.352e-4 | 16.0514 | 4.593e-4 | 16.0531 | 4.75e-4 |
| $q_s C_s$ | 11.0832 | 3.927e-4 | 11.1051 | 4.050e-4 | 11.1358 | 4.098e-4 |
| $\tau_1$ | 6.475e-3 | 1.181e-4 | 0.0116 | 1.205e-4 | 0.0203 | 1.224e-3 |
| $\tau_2$ | 2.642e-3 | 1.547e-4 | 0.0102 | 1.556e-4 | 0.0174 | 1.541e-4 |
| $\tau_3$ | 0.9907 | 3.025e-4 | 0.9778 | 3.066e-4 | 0.9618 | 3.070e-4 |
| $\gamma_1$ | 0.1711 | 1.536e-5 | 0.1732 | 1.542e-5 | 0.1748 | 1.516e-5 |
| $\gamma_2$ | 0.6979 | 2.524e-4 | 0.6871 | 2.565e-4 | 0.6744 | 2.601e-4 |

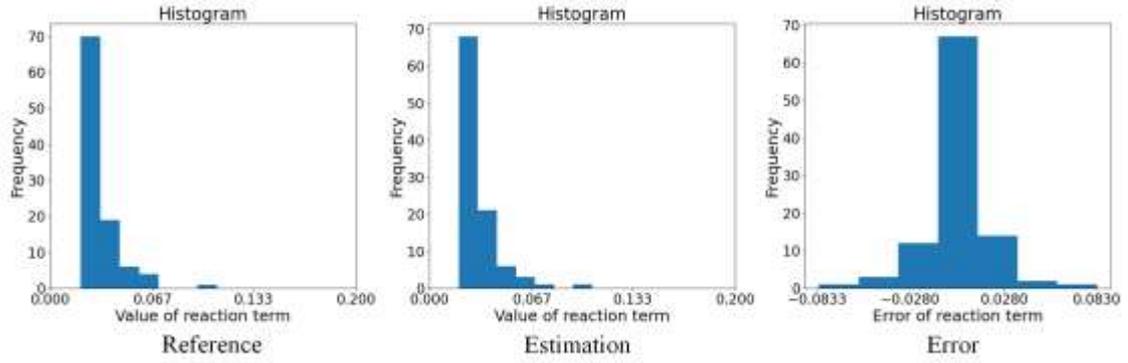

**Figure 19.** Retardation effect of the reference and estimated sorption terms.

### 3.3 Influence of strong nonlinearity and information volume on data assimilation results

In this section, influences of strong nonlinearity and information volume on data assimilation results are discussed. In order to investigate the case with increased nonlinearity and decreased information volume, hydraulic conductivity fields and source parameters are simultaneously estimated by using only contaminant observation data. Observations of the hydraulic head are assumed to be lacking, and Freundlich sorption is chosen as the sorption type in the studied case. Initialized parameters ***m*** in the IES method include $K$ fields and source parameters. TgU-net-h and TgU-net-C surrogate models are connected in series. Firstly, $h$ fields are predicted through TgU-net-h, and the calculated seepage fields are fed into TgU-net-C to generate corresponding contaminant fields. Then, the IES method can be implemented to estimate the hydraulic $K$ fields and source parameters by only using observation data from $C$ fields. No sparsity-promoting techniques are employed.

It is worth mentioning that configurations of the IES method in this section are the same as those in section 3.2. The RMSE of inversed of ln$K$ fields with different noise levels is presented in **Table 10**, and the estimated source parameters are displayed in **Table C1** in Appendix C. By comparison with the results in section 3.2, it can be seen that the data assimilation results using both $h$ field data and $C$ field data are closer to the reference value than those only using contaminant field data. Unknown conductivity fields yield uncertainty of flow fields, which may increase the difficulty of data assimilation. More data or more information are beneficial to reduce the posterior uncertainty of data assimilation results. Furthermore, predicting the contaminant fields through conductivity fields will lead to stronger nonlinearity. The increased nonlinearity of the system may also result in the increased posterior uncertainty in the inversion process. Moreover, prediction error may be transmitted between the two surrogate models, further rendering inaccurate results.

**Table 10**. RMSE of the mean of estimated ln$K$ with different noise levels (Freundlich sorption).

|  | Layer 1 | Layer 2 | Layer 3 | Layer 4 | Layer 5 |
|---|---|---|---|---|---|
|  | 0.1850 | 0.1529 | 0.1262 | 0.1075 | 0.1407 |
| Clean | Layer 6 | Layer 7 | Layer 8 | Layer 9 | Layer 10 |
| data | 0.1551 | 0.1331 | 0.1400 | 0.1157 | 0.1016 |
|  | Layer 11 | Layer 12 | Layer 13 | Layer 14 | Layer 15 |
|  | 0.1139 | 0.1431 | 0.1738 | 0.1923 | 0.1305 |
|  | Layer 1 | Layer 2 | Layer 3 | Layer 4 | Layer 5 |
|  | 0.2088 | 0.2433 | 0.3048 | 0.1987 | 0.2428 |
| 1% | Layer 6 | Layer 7 | Layer 8 | Layer 9 | Layer 10 |
| noise | 0.2335 | 0.1748 | 0.2000 | 0.1639 | 0.1719 |
|  | Layer 11 | Layer 12 | Layer 13 | Layer 14 | Layer 15 |
|  | 0.1819 | 0.1844 | 0.2186 | 0.2087 | 0.1616 |
|  | Layer 1 | Layer 2 | Layer 3 | Layer 4 | Layer 5 |
|  | 0.2101 | 0.2484 | 0.3026 | 0.1935 | 0.2502 |
| 5% | Layer 6 | Layer 7 | Layer 8 | Layer 9 | Layer 10 |
| noise | 0.2496 | 0.1837 | 0.2049 | 0.1671 | 0.1788 |
|  | Layer 11 | Layer 12 | Layer 13 | Layer 14 | Layer 15 |
|  | 0.1848 | 0.1828 | 0.2493 | 0.2384 | 0.1750 |
|  | Layer 1 | Layer 2 | Layer 3 | Layer 4 | Layer 5 |
|  | 0.2273 | 0.2664 | 0.4020 | 0.2415 | 0.3829 |
| 10% | Layer 6 | Layer 7 | Layer 8 | Layer 9 | Layer 10 |
| noise | 0.4094 | 0.2552 | 0.2198 | 0.1866 | 0.2192 |
|  | Layer 11 | Layer 12 | Layer 13 | Layer 14 | Layer 15 |
|  | 0.1989 | 0.2049 | 0.4372 | 0.3593 | 0.2149 |

## 4. Discussion and Conclusions

In this work, the TgU-net model is proposed for surrogate modeling of contaminant transport problems. Since training data are usually hard to obtain, TgU-net incorporates governing equation constraints into the loss function of the traditional U-net to regulate the training of the constructed neural networks, which can increase prediction accuracy and decrease data requirements. In addition, there is no need to parameterize the high-dimensional spatial field due to the convolutional structure that is used in U-net. Different from traditional methods, in which one model corresponds to one governing equation, this proposed TgU-net can deal with physical problems with uncertain processes that can be descried by multiple potential models through only one surrogate model. In the considered 3D contaminant convection-dispersion problem, sorption is considered to be a potential process, and three sorption types (i.e., linear, Freundlich, and Langmuir sorption) are integrated into one comprehensive underlying governing equation by assigning indicators. By evaluating RMSE and $R^2$ score, TgU-net demonstrates satisfactory fitting, generalizability, and extrapolability. The constructed TgU-net surrogate models can provide accurate predictions, even when dealing with configurations that are out of training range, such as hydraulic conductivity fields with higher mean and variance.

For physical problems that involve multiple processes, the dominant processes and the proper empirical model for a specific process may be unclear. In order to identify the governing equation of this type of problem, methods, such as model selection based on symbolic regression and information criterion, usually require a large number of model executions, resulting in large computational cost. To solve this problem, the proposed TgU-net, which is trained by the comprehensive governing equation, can be combined with IES for governing equation identification of this type of problem. For the investigated contaminant transport problem, unknown hydraulic parameters, contaminant source information, and governing sorption process are identified simultaneously. The estimated sorption indicators can determine the existing sorption type. The convergence of the estimation to the reference values indicates the validity of this proposed method. Influences of inevitable data noise, strong nonlinearity, and information volume are also explored. Compared to the case in which both hydraulic head data and contaminant field data are used, the results that only use contaminant field data are worse. The variance of estimated *K* fields and source parameters can be larger due to the existence of noise.

This proposed TgU-net based method provides a novel way for identifying physical problems that involve multiple processes. Although the proposed framework is only tested by using a contaminant transport problem in this work, it may be effectively extended to a wider range of problems.

**Acknowledgements**

This work is partially funded by the Shenzhen Key Laboratory of Natural Gas Hydrates (Grant No. ZDSYS20200421111201738) and the SUSTech - Qingdao New Energy Technology Research Institute.

**References**

Brunton, S. L., Proctor, J. L., Kutz, J. N., & Bialek, W. (2016). Discovering governing equations from data by sparse identification of nonlinear dynamical systems. *Proceedings of The National Academy of Sciences - PNAS, 113*(15), 3932-3937. doi:10.1073/pnas.1517384113.

Cao, T., Zeng, X., Wu, J., Wang, D., Sun, Y., Zhu, X., Long, Y., et al. (2019). Groundwater contaminant source identification via Bayesian model selection and uncertainty quantification. *Hydrogeology Journal, 27*(8), 2907-2918. doi:10.1007/s10040-019-02055-3.

Chang, H., Liao, Q., & Zhang, D. (2017). Surrogate model based iterative ensemble smoother for subsurface flow data assimilation. *Advances in Water Resources*, 100, 96-108. doi:10.1016/j.advwatres.2016.12.001.

Chang, H., & Zhang, D. (2019). Identification of physical processes via combined data-driven and data-assimilation methods. *Journal of Computational Physics, 393*, 337-350. doi:10.1016/j.jcp.2019.05.008.

Chen, J., Viquerat, J., & Hachem, E. J. a. C. P. (2020). U-net architectures for fast prediction of incompressible laminar flows. arXiv preprint arXiv: 1910.13532.

Chen, Y., & Oliver, D. S. (2013). Levenberg–Marquardt forms of the iterative ensemble smoother for efficient history matching and uncertainty quantification. *Computational Geosciences, 17*(4), 689-703. doi:10.1007/s10596-013-9351-5.

Chun-Yu, G., Yi-Wei, F., Yang, H., Peng, X., & Yun-Fei, K. (2021). Deep-learning-based liquid extraction algorithm for particle image velocimetry in two-phase flow experiments of an object entering water. *Applied Ocean Research, 108*, 102526. doi:10.1016/j.apor.2021.102526.

Dolz, J., Ben Ayed, I., & Desrosiers, C. (2019). Dense multi-path U-net for ischemic stroke lesion segmentation in multiple image modalities. *Brainlesion: Glioma, Multiple Sclerosis, Stroke and Traumatic Brain Injuries* (pp. 271-282). Springer International Publishing. doi:10.1007/978-3-030-11723-8_27.

Fetter, C. W. (1999). Contaminant hydrogeology, 2nd ed., Prentice Hall.

He, T., Wang, N., & Zhang, D. (2021). Theory-guided full convolutional neural network: An efficient surrogate model for inverse problems in subsurface contaminant transport. *Advances in Water Resources, 157*, 104051. doi:10.1016/j.advwatres.2021.104051.

Imambi, S., Prakash, K. B., & Kanagachidambaresan, G. R. (2021). Pytorch. *Programming with TensorFlow* (pp. 87-104). Springer International Publishing. doi:10.1007/978-3-030-57077-4_10.

Jiang, Z., Tahmasebi, P., & Mao, Z. (2021). Deep residual U-net convolution neural networks with autoregressive strategy for fluid flow predictions in large-scale geosystems. *Advances in Water Resources, 150*, 103878. doi:10.1016/j.advwatres.2021.103878.

Kingma, D. P., & Ba, J. L. (2015). Adam: A method for stochastic optimization. Paper presented at the International Conference on Learning Representations.

Kuha, J. (2004). AIC and BIC: Comparisons of assumptions and performance. *Sociological Methods and Research, 33*(2), 188-229. doi:10.1177/0049124103262065.

Lakshmi, M. V. S., Saisreeja, P. L., Chandana, L., Mounika, P., & U, P. (2021). A LeakyReLU based effective brain MRI segmentation using U-NET. Paper presented at the 1251-1256. doi:10.1109/ICOEI51242.2021.9453079.

Le, Q. T., & Ooi, C. (2021). Surrogate modeling of fluid dynamics with a multigrid inspired neural network architecture. *Machine Learning with Applications, 6*. doi:10.1016/j.mlwa.2021.100176.

Lee, J.-Y., & Park, J. (2021). Deep regression network-assisted efficient streamline generation method. *IEEE Access, 9*, 111704-111717. doi:10.1109/ACCESS.2021.3100127.

Loshchilov, I., & Hutter, F. (2017). Fixing weight decay regularization in Adam. arXiv preprint arXiv: 1711.05101.

Mangan, N. M., Kutz, J. N., Brunton, S. L., & Proctor, J. L. (2017). Model selection for dynamical systems via sparse regression and information criteria. *Proceedings of the Royal Society. A, Mathematical, Physical, and Engineering Sciences, 473*(2204), 20170009-20170009. doi:10.1098/rspa.2017.0009.

Mo, S., Zabaras, N., Shi, X., & Wu, J. (2019a). Deep autoregressive neural networks for high-dimensional inverse problems in groundwater contaminant source identification. *Water Resources Research, 55*(5), 3856-3881.

Mo, S., Zhu, Y., Zabaras, N., Shi, X., & Wu, J. (2019b). Deep convolutional encoder-decoder networks for uncertainty quantification of dynamic multiphase flow in heterogeneous media. *Water Resources Research, 55*, 703–728. doi:10.1029/2018WR023528.

Newell, A., Yang, K., & Deng, J. (2016). Stacked hourglass networks for human pose estimation. *Computer vision – ECCV 2016* (pp. 483-499). Springer International Publishing. doi:10.1007/978-3-319-46484-8_29.

Oliver, D. S., Reynolds, A. C., & Liu, N. (2008). Inverse theory for petroleum reservoir characterization and history matching. Cambridge University Press. doi:10.1017/CBO9780511535642.

Raissi, M., Perdikaris, P., & Karniadakis, G. E. (2019). Physics-informed neural networks: A deep learning framework for solving forward and inverse problems involving nonlinear partial differential equations. *Journal of Computational Physics, 378*, 686-707. doi:10.1016/j.jcp.2018.10.045.

Ronneberger, O., Fischer, P., & Brox, T. (2015). U-net: Convolutional networks for biomedical image segmentation. (pp. 234-241). Springer International Publishing. doi:10.1007/978-3-319-24574-4_28.

Schaeffer, H. (2017). Learning partial differential equations via data discovery and sparse optimization. *Proceedings of The Royal Society. A, Mathematical, Physical, and Engineering Sciences, 473*(2197), 20160446-20160446. doi:10.1098/rspa.2016.0446.

Schoeniger, A., Woehling, T., Samaniego, L., & Nowak, W. (2014). Model selection on solid ground: Rigorous comparison of nine ways to evaluate Bayesian model evidence. *Water Resources Research, 50*(12), 9484-9513. doi:10.1002/2014WR016062.

Shelhamer, E., Long, J., & Darrell, T. (2017). Fully convolutional networks for semantic segmentation. *IEEE Transactions on Pattern Analysis and Machine Intelligence, 39*(4), 640-651. doi:10.1109/TPAMI.2016.2572683.

Srivastava, D., & Singh, R. M. (2015). Groundwater system modeling for simultaneous identification of pollution sources and parameters with uncertainty characterization. *Water Resources Management, 29*(13), 4607-4627. doi:10.1007/s11269-015-1078-8.

Tang, M., Liu, Y., & Durlofsky, L. J. (2020). A deep-learning-based surrogate model for data assimilation in dynamic subsurface flow problems. *Journal of Computational Physics, 413*, 109456. doi:10.1016/j.jcp.2020.109456.

Tang, Z., Peng, X., Geng, S., Zhu, Y., & Metaxas, D. N. (2018). CU-Net: Coupled U-Nets. Paper presented at the BMVC.

Troldborg, M., Nowak, W., Tuxen, N., Bjerg, P. L., Helmig, R., & Binning, P. J. (2010). Uncertainty evaluation of mass discharge estimates from a contaminated site using a fully bayesian framework. *Water Resources Research, 46*(12), n/a. doi:10.1029/2010WR009227.

Wang, N., Chang, H., & Zhang, D. (2021a). Efficient uncertainty quantification for dynamic subsurface flow with surrogate by theory-guided neural network. *Computer Methods in Applied Mechanics and Engineering, 373.* doi:10.1016/j.cma.2020.113492.

Wang, N., Chang, H., & Zhang, D. (2021b). Efficient uncertainty quantification and data assimilation via theory-guided convolutional neural network. Paper presented at the SPE Reservoir Simulation Conference, Galveston, Texas, USA. Society of Petroleum Engineers.

Wang, N., Zhang, D., Chang, H., & Li, H. (2020a). Deep learning of subsurface flow via theory-guided neural network. *Journal of Hydrology (Amsterdam), 584*, 124700. doi:10.1016/j.jhydrol.2020.124700.

Wang, Y. D., Chung, T., Armstrong, R. T., & Mostaghimi, P. (2020b). Ml-lbm: Machine learning aided flow simulation in porous media. arXiv preprint arXiv: 2004.11675.

Wu, H., Fang, W., Kang, Q., Tao, W., Qiao, R., & Los Alamos National Lab. (LANL), Los Alamos, NM (United States). (2019). Predicting effective diffusivity of porous media from images by deep learning. *Scientific Reports*, 9(1), 20387-12. doi:10.1038/s41598-019-56309-x.

Xu, H., Chang, H., & Zhang, D. (2020). DLGA-PDE: Discovery of PDEs with incomplete candidate library via combination of deep learning and genetic

algorithm. *Journal of Computational Physics, 418*, 109584. doi:10.1016/j.jcp.2020.109584.

Xu, R., Wang, N., & Zhang, D. (2021). Solution of diffusivity equations with local sources/sinks and surrogate modeling using weak form theory-guided neural network. *Advances in Water Resources, 153*, 103941. doi:10.1016/j.advwatres.2021.103941.

Xu, T., & Gómez-Hernández, J. J. (2018). Simultaneous identification of a contaminant source and hydraulic conductivity via the restart normal-score ensemble Kalman filter. *Advances in Water Resources, 112*, 106-123. doi: 10.1016/j.advwatres.2017.12.011.

Yang, L., Zhang, D., Karniadakis, G. E. M., & Brown Univ., Providence, RI (United States). (2020). Physics-informed generative adversarial networks for stochastic differential equations. *SIAM Journal on Scientific Computing, 42*(1), A292-A317. doi:10.1137/18M1225409.

Ye, M., Meyer, P. D., Neuman, S. P., & Pacific Northwest National Lab. (PNNL), Richland, WA (United States). (2008). On model selection criteria in multimodel analysis. *Water Resources Research, 44*(3), W03428-n/a. doi:10.1029/2008WR006803.

Ying, S., Zhang, J., Zeng, L., Shi, J., & Wu, L. (2017). Bayesian inference for kinetic models of biotransformation using a generalized rate equation. *The Science of The Total Environment, 590-591*, 287-296. doi:10.1016/j.scitotenv.2017.03.003.

Zhang, D., & Lu, Z. (2004). An efficient, high-order perturbation approach for flow in random porous media via Karhunen–Loève and polynomial expansions. *Journal of Computational Physics, 194*(2), 773-794. doi:10.1016/j.jcp.2003.09.015

Zhang, J., Zeng, L., Chen, C., Chen, D., & Wu, L. (2015). Efficient Bayesian experimental design for contaminant source identification. *Water Resources Research, 51*(1), 576-598. doi: 10.1002/2014WR015740.

Zheng, C. , & Wang, P. P. . (1999). Mt3dms: A modular three-dimensional multispecies transport model for simulation of advection, dispersion, and chemical reactions of contaminants in groundwater systems; documentation and user's guide. *Ajr American Journal of Roentgenology, 169*(4), 1196-7.

Zhou, Z., & Tartakovsky, D. M. (2021). Markov chain Monte Carlo with neural network surrogates: Application to contaminant source identification. *Stochastic Environmental Research and Risk Assessment, 35*(3), 639-651. doi:10.1007/s00477-020-01888-9.

**Appendix**

**Appendix A**

In this section, fitting ability, generalizability, and extrapolability of the constructed TgU-net-based surrogate models are tested with sparse data points. In addition, the corresponding traditional U-nets are trained in the same configurations as comparisons. Predicted *h* fields of U-net-h and TgU-net-h are presented in **Fig. A1** and **Fig. A2** in Appendix A, respectively. RMSEs of these points predicted by U-net and TgU-net are given in **Table A1** in Appendix A.

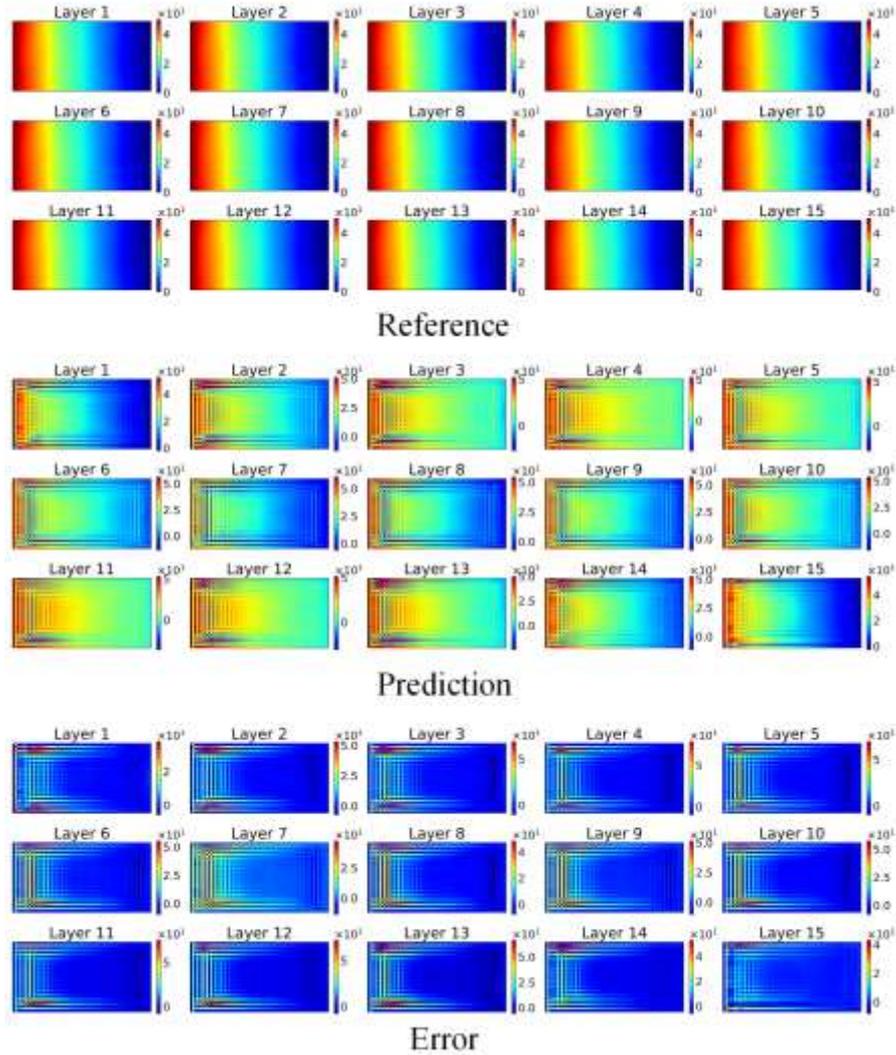

**Figure A1.** Prediction and error of U-net-h when $\overline{lnK(\mathbf{x})}$ is set to be 1 and variance of *lnK* is 0.5.

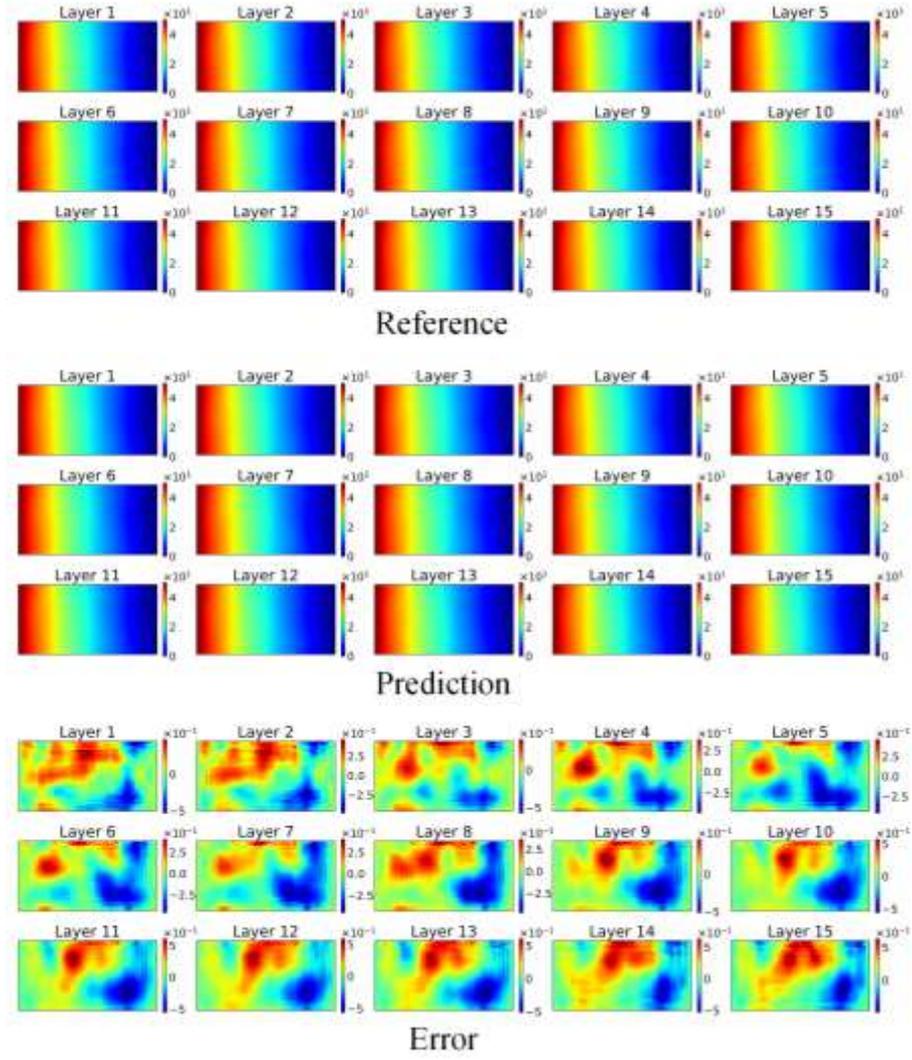

**Figure A2.** Prediction and error of TgU-net-h when $\overline{lnK(\mathbf{x})}$ is set to be 1 and variance of *lnK* is 0.5.

**Table A1**. RMSE of points where observation data are extracted.

| | Layer 1 | Layer 2 | Layer 3 | Layer 4 | Layer 5 |
|---|---|---|---|---|---|
| U-net-h | 0.2074 | 0.2429 | 0.2776 | 0.2657 | 0.2593 |
| | Layer 6 | Layer 7 | Layer 8 | Layer 9 | Layer 10 |
| | 0.2462 | 0.2365 | 0.2318 | 0.2365 | 0.2514 |
| | Layer 11 | Layer 12 | Layer 13 | Layer 14 | Layer 15 |
| | 0.2601 | 0.2838 | 0.2820 | 0.2703 | 0.2837 |
| TgU-net-h | Layer 1 | Layer 2 | Layer 3 | Layer 4 | Layer 5 |
| | 0.1932 | 0.1888 | 0.1708 | 0.1715 | 0.1765 |
| | Layer 6 | Layer 7 | Layer 8 | Layer 9 | Layer 10 |
| | 0.1735 | 0.1760 | 0.1848 | 0.2008 | 0.2162 |
| | Layer 11 | Layer 12 | Layer 13 | Layer 14 | Layer 15 |
| | 0.2259 | 0.2232 | 0.2326 | 0.2446 | 0.2447 |

**Appendix B**

In this section, the trained TgU-net-based surrogate models are employed to process forward predictions, and are combined with the IES method for inverse modeling. Firstly, observation data of hydraulic head fields are used to estimate $K$ fields. Then, the velocity fields from the estimated $K$ fields are inputted into TgU-net-C to estimate unknown source parameters and physical processes according to observation data of contaminant concentration. **Fig. B1** shows prediction and error of the hydraulic head field of the estimated ln$K$ field in the IES method. Comparison of convergence of initialized and estimated parameters of linear and Freundlich sorption is shown in **Fig. B2** and **B3** in Appendix B, respectively. Convergence of contaminant parameters of linear and Freundlich sorption are presented in **Table B1** and **B2** in Appendix B, respectively. **Table B3** compares the convergence results of parameters by utilizing different sparsity-promoting methods under 10% noise.

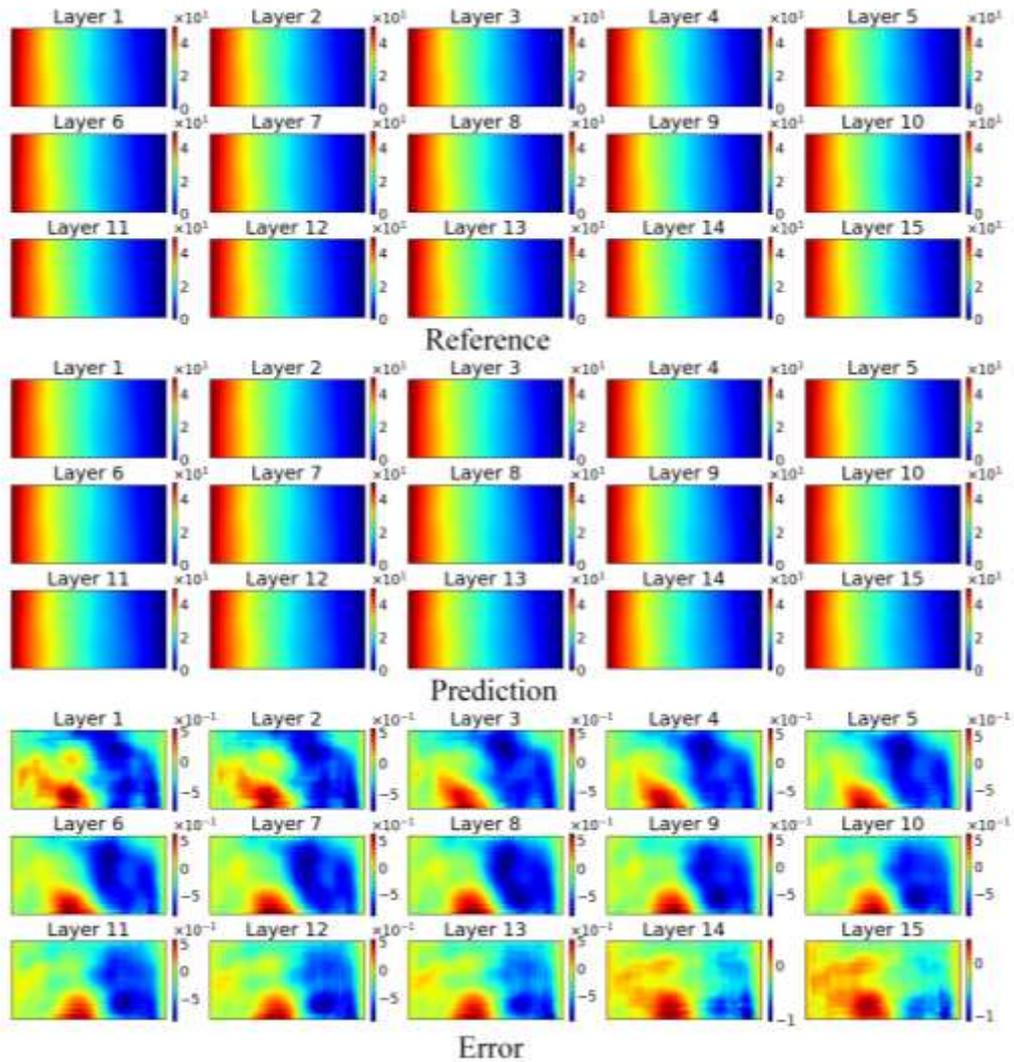

**Figure B1.** Prediction and error of the hydraulic head field of estimated ln$K$ field in the IES method.

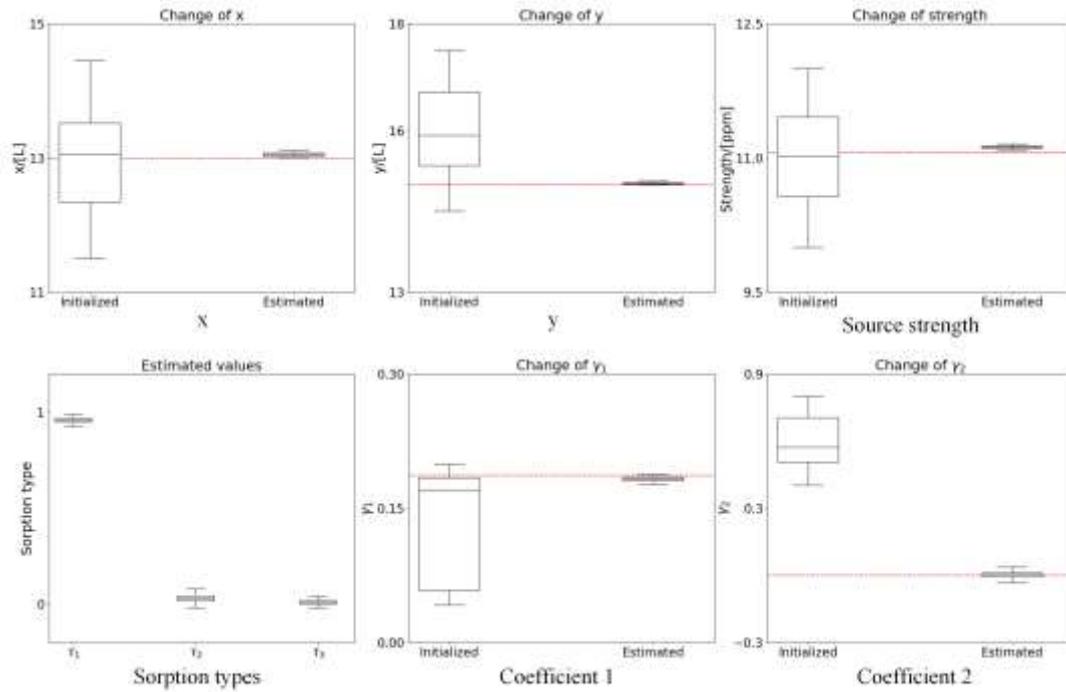

**Figure B2.** Comparison of convergence of initialized and estimated parameters (linear sorption).

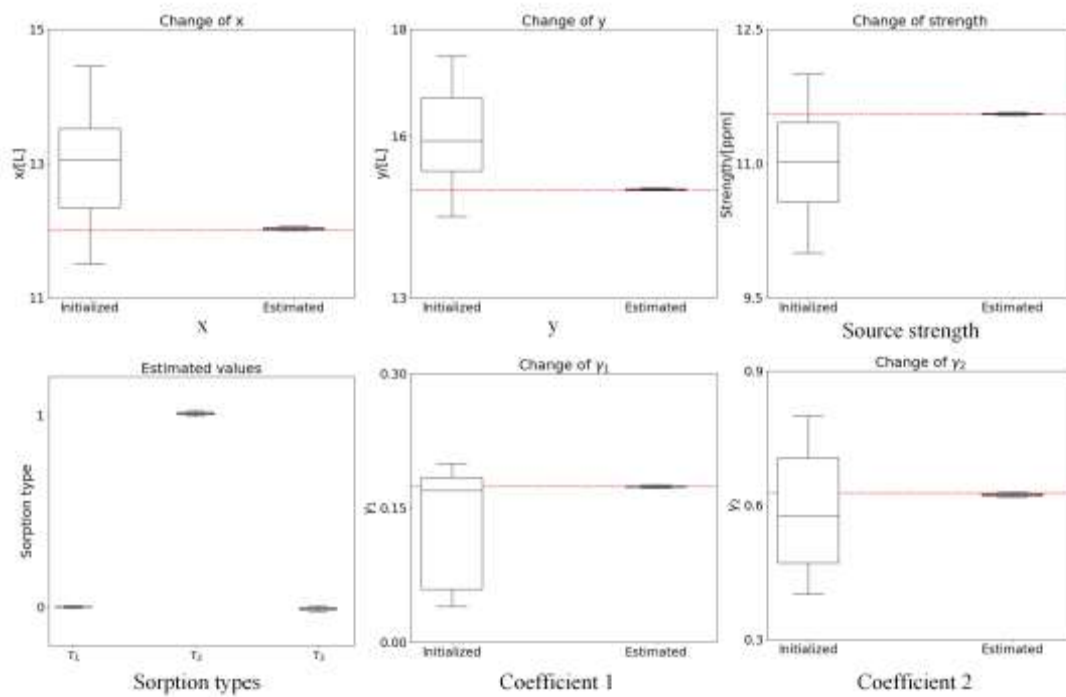

**Figure B3.** Comparison of convergence of initialized and estimated parameters (Freundlich sorption).

Table B1. Convergence of parameters under different noise levels (linear sorption).

| Parameters | Reference | Clean data | | 1% noise | | 5% noise | | 10% noise | |
|---|---|---|---|---|---|---|---|---|---|
| | | Mean | Variance | Mean | Variance | Mean | Variance | Mean | Variance |
| $x$ | 13 | 13.0503 | 7.542e-4 | 13.0533 | 8.051e-4 | 13.0495 | 7.785e-4 | 13.0515 | 7.425e-3 |
| $y$ | 15 | 15.0302 | 3.532e-4 | 15.0307 | 5.006e-4 | 15.0315 | 3.562e-4 | 15.1291 | 3.728e-4 |
| $q_s C_s$ | 11.0652 | 11.0782 | 2.936e-4 | 11.1192 | 3.165e-4 | 11.1677 | 2.873e-4 | 11.2586 | 5.293e-4 |
| $\tau_1$ | 1 | 0.9862 | 3.788e-4 | 0.9569 | 4.085e-4 | 0.8642 | 3.679e-4 | 0.7544 | 4.289e-4 |
| $\tau_2$ | 0 | 2.153e-2 | 7.112e-4 | 3.576e-2 | 7.578e-4 | 7.867e-2 | 7.186e-4 | 0.1515 | 9.131e-4 |
| $\tau_3$ | 0 | 1.184e-2 | 3.669e-4 | 7.664e-3 | 4.053e-4 | 5.715e-2 | 3.53e-4 | 9.368e-2 | 3.739e-4 |
| $\gamma_1$ | 0.1865 | 0.1854 | 8.314e-6 | 0.1828 | 8.417e-6 | 0.1869 | 8.551e-6 | 0.1898 | 9.172e-6 |
| $\gamma_2$ | 0 | 8.865e-3 | 4.808e-4 | 4.999e-3 | 5.111e-4 | 1.141e-3 | 4.452e-4 | 1.519e-2 | 4.714e-4 |

Table B2. Convergence of parameters under different noise levels (Freundlich sorption).

| Parameters | Reference | Clean data | | 1% noise | | 5% noise | | 10% noise | |
|---|---|---|---|---|---|---|---|---|---|
| | | Mean | Variance | Mean | Variance | Mean | Variance | Mean | Variance |
| $x$ | 12 | 12.0166 | 1.167e-4 | 12.0319 | 2.806e-4 | 12.042 | 4.984e-4 | 12.0461 | 5.189e-4 |
| $y$ | 15 | 15.0096 | 4.361e-5 | 15.0199 | 1.527e-4 | 15.040 | 3.576e-4 | 15.0302 | 4.429e-4 |
| $q_s C_s$ | 11.5450 | 11.5482 | 2.569e-6 | 11.5538 | 3.835e-5 | 11.5619 | 2.610e-4 | 11.4517 | 3.883e-4 |
| $\tau_1$ | 0 | 1.249e-4 | 1.190e-7 | 4.186e-4 | 4.910e-6 | 9.212e-4 | 4.352e-5 | 2.031e-2 | 4.137e-5 |
| $\tau_2$ | 1 | 0.9973 | 2.825e-6 | 0.9898 | 3.941e-5 | 0.9854 | 3.773e-5 | 0.8921 | 6.496e-4 |
| $\tau_3$ | 0 | 2.453e-4 | 2.463e-6 | 9.853e-3 | 3.268e-5 | 5.431e-3 | 3.774e-4 | 0.1684 | 5.918e-4 |
| $\gamma_1$ | 0.1738 | 0.1746 | 5.497e-8 | 0.1741 | 6.619e-7 | 0.1767 | 4.224e-6 | 0.1796 | 5.257e-6 |
| $\gamma_2$ | 0.6272 | 0.6269 | 6.808e-7 | 0.6234 | 8.135e-6 | 0.6307 | 5.267e-5 | 0.6794 | 7.423e-5 |

Table B3. Comparison of parameters convergence of different methods under 10% noise (Freundlich sorption).

| Parameters | Reference | Normal | | Signal | | LASSO ($\alpha = 0.05$) | | LASSO ($\alpha = 0.1$) | |
|---|---|---|---|---|---|---|---|---|---|
| | | Mean | Variance | Mean | Variance | Mean | Variance | Mean | Variance |
| $x$ | 12 | 12.0461 | 5.189e-4 | 12.0440 | 4.646e-4 | 12.0170 | 2.509e-4 | 12.0217 | 3.578e-4 |
| $y$ | 15 | 15.0302 | 4.429e-4 | 15.0338 | 5.947e-4 | 15.0881 | 6.689e-4 | 15.1169 | 6.960e-4 |
| $q_s C_s$ | 11.5450 | 11.4517 | 3.883e-4 | 11.6044 | 2.697e-4 | 11.6132 | 4.847e-5 | 11.6138 | 5.470e-5 |
| $\tau_1$ | 0 | 2.031e-2 | 4.137e-5 | 0 | 0 | 1.472e-2 | 7.657e-6 | 1.843e-3 | 7.018e-6 |
| $\tau_2$ | 1 | 0.8921 | 6.496e-4 | 1 | 0 | 0.9131 | 4.320e-5 | 0.9842 | 4.515e-5 |
| $\tau_3$ | 0 | 0.1684 | 5.918e-4 | 0 | 0 | 7.22e-2 | 2.946e-5 | 1.397e-2 | 2.959e-5 |
| $\gamma_1$ | 0.1738 | 0.1796 | 5.257e-6 | 0.1785 | 5.818e-6 | 0.1866 | 4.168e-7 | 0.1823 | 4.972e-7 |
| $\gamma_2$ | 0.6272 | 0.6794 | 7.423e-5 | 0.6254 | 2.818e-5 | 0.6227 | 6.857e-6 | 0.6289 | 6.322e-6 |

**Appendix C**

In this section, the influences of strong nonlinearity and information volume on data assimilation results are discussed. Estimated source parameters are displayed in **Table C1**.

Table C1. Parameters convergence of the IES method only using data from $C$ fields under different noise levels (Freundlich sorption).

| Parameters | Reference | Clean data | | 1% noise | | 5% noise | | 10% noise | |
|---|---|---|---|---|---|---|---|---|---|
| | | Mean | Variance | Mean | Variance | Mean | Variance | Mean | Variance |
| $x$ | 12 | 12.1092 | 2.354e-4 | 12.1346 | 1.305e-3 | 12.1391 | 1.373e-3 | 12.1614 | 1.461e-3 |
| $y$ | 15 | 15.0872 | 1.756e-5 | 15.0927 | 6.677e-4 | 15.1308 | 7.061e-4 | 15.0989 | 6.915e-4 |
| $q_s C_s$ | 11.5450 | 11.4812 | 1.962e-3 | 11.3412 | 1.533e-2 | 11.3756 | 1.632e-2 | 11.1364 | 1.646e-2 |
| $\tau_1$ | 0 | 7.812e-3 | 4.196e-4 | 6.708e-4 | 2.675e-3 | 6.723e-3 | 2.725e-3 | 1.172e-2 | 2.692e-3 |
| $\tau_2$ | 1 | 0.9485 | 1.337e-3 | 0.9199 | 5.877e-3 | 0.9058 | 6.294e-3 | 0.7749 | 7.496e-3 |
| $\tau_3$ | 0 | 5.042e-2 | 6.467e-6 | 6.294e-2 | 4.159e-3 | 7.272e-2 | 4.394e-3 | 0.2077 | 5.297e-3 |
| $\gamma_1$ | 0.1738 | 0.1785 | 4.886e-6 | 0.1780 | 2.351e-5 | 0.1789 | 2.400e-5 | 0.1842 | 2.594e-5 |
| $\gamma_2$ | 0.6272 | 0.6351 | 2.765e-4 | 0.6036 | 1.754e-3 | 0.6044 | 1.740e-3 | 0.6353 | 1.579e-3 |